\newcommand*\myhrulefill{%
\leavevmode\leaders\hrule depth-2pt height 2.4pt\hfill\kern0pt
}

\documentclass[useAMS,usenatbib]{mn2e}

\usepackage{graphicx}
\usepackage{url}
\usepackage{bm}

\newcommand{\chapin}{C09a}
\newcommand{\patanchon}{P09}

\def\lsim{\mathrel{\lower2.5pt\vbox{\lineskip=0pt\baselineskip=0pt
           \hbox{$<$}\hbox{$\sim$}}}}
\def\gsim{\mathrel{\lower2.5pt\vbox{\lineskip=0pt\baselineskip=0pt
           \hbox{$>$}\hbox{$\sim$}}}}

\title[BLAST: Evolution of FIR Luminosity Function]{A Monte Carlo Approach to Evolution of the Far-Infrared Luminosity Function with BLAST}
\author[G. Marsden et al.]{Gaelen~Marsden,$^{1}$\thanks{E-mail: gmarsden@phas.ubc.ca} Edward~L.~Chapin,$^{1}$
Mark~Halpern,$^{1}$
Guillaume~Patanchon,$^{2}$\newauthor
Douglas~Scott,$^{1}$
Matthew~D.~P.~Truch,${^3}$
Elisabetta~Valiante,${^1}$
Marco~P.~Viero,$^{4,5}$\newauthor
Donald~V.~Wiebe$^{1}$
\\
$^{1}$Department of Physics \& Astronomy, University of
British Columbia, 6224 Agricultural Road, Vancouver, BC V6T~1Z1,
Canada\\
$^2$Universit\'{e} Paris Diderot, Laboratoire APC, 10, rue Alice
Domon et L\'{e}onie Duquet 75205 Paris, France\\
$^3$Department of Physics \& Astronomy, University of
  Pennsylvania, 209 South 33rd Street, Philadelphia, PA, 19104, USA\\
$^4$California Institute of Technology, 1200 E. California Blvd.,
Pasadena, CA 91125, USA\\
$^5$Department of Astronomy \& Astrophysics, University of
Toronto, 50 St. George Street, Toronto, ON M5S 3H4, Canada
}

\begin{document}

\pagerange{\pageref{firstpage}--\pageref{lastpage}} \pubyear{2010}

\maketitle

\label{firstpage}

\begin{abstract}
We constrain the evolution of the rest-frame far-infrared (FIR)
luminosity function out to high redshift, by combining several pieces
of complementary information provided by the deep Balloon-borne
Large-Aperture Submillimeter Telescope surveys at 250, 350 and
500\,\micron, as well as other FIR and millimetre data. Unlike most
other phenomenological models, we characterise the uncertainties in
our fitted parameters using Monte Carlo Markov Chains. We use a
bivariate local luminosity function that depends only on FIR
luminosity and 60-to-100\,\micron\ colour, along with a single library
of galaxy spectral energy distributions indexed by colour, and apply
simple luminosity and density evolution. We use the surface density of
sources, Cosmic Infrared Background (CIB) measurements and redshift
distributions of bright sources, for which identifications have been
made, to constrain this model. The precise evolution of the FIR
luminosity function across this crucial range has eluded studies at
longer wavelengths (e.g., using SCUBA and MAMBO) and at shorter
wavelengths (e.g., with \textsl{Spitzer\/}), and should provide a key
piece of information required for the study of galaxy evolution. Our
adoption of Monte Carlo methods enables us not only to find the
best-fit evolution model, but also to explore correlations between the
fitted parameters. Our model-fitting approach allows us to focus on
sources of tension coming from the combination of data-sets.  We
specifically find that our choice of parameterisation has difficulty
fitting the combination of CIB measurements and redshift distribution
of sources near $1\,$mm.  Existing and future data sets will be able
to dramatically improve the fits, as well as break strong degeneracies
among the models.  Two particular examples that we find to be crucial
are: obtaining robust information on redshift distributions; and
placing tighter constraints on the range of spectral shapes for low
luminosity $(L_\mathrm{FIR}<10^{10}\,\mathrm{L}_\odot)$ sources.
\end{abstract}


\section{Introduction}
\label{sec:intro}

It is now known that a significant fraction of the total light
produced by stars and active galactic nuclei (AGN) throughout cosmic
history is absorbed by dust and re-radiated thermally at much longer
wavelengths. This light was first observed at low angular resolution
by the \textsl{COBE\/} satellite as the diffuse Cosmic Infrared
Background \citep[CIB,][]{puget1996,fixsen1998}. Over the last decade
it has been largely resolved into point sources at wavelengths
${\sim}\,100$--1000\,\micron, demonstrating that it is predominantly
produced by individual galaxies
\citep{dole2006,pope2007,serjeant2008,marsden2009,pascale2009}. Surveys
with the \textsl{IRAS\/}, \textsl{ISO\/} and \textsl{Spitzer\/}
satellites show that most of the shorter-wavelength light is produced
by galaxies at redshifts $z \la 1$, while ground-based submillimetre
(submm) surveys have found most of the longer-wavelength light to be
produced by more distant objects. Recent surveys by the Balloon-borne
Large Aperture Submillimeter Telescope (BLAST) at 250, 350 and
500\,\micron, a precursor to the \textsl{Herschel\/}/SPIRE surveys
that are now well underway, have shown that the transition from low to
high redshifts as one observes at longer wavelengths occurs gradually
across the 250--500\,\micron\ band.

The fact that there is a transition to higher-redshift sources
observed at longer wavelengths is not a surprise. Many groups have
predicted this general behaviour using simple parameterised models for
the evolution of local far-infrared (FIR) and submm galaxy luminosity
functions. The data typically fit at these wavelengths include the
surface density of sources as a function of brightness (source counts)
and redshift information, when available
\citep[e.g.,][]{blain1993,guiderdoni1997,blain1999,chary2001,malkan2001,rowan-robinson2001,lagache2003,dole2003,lagache2004,leborgne2009,valiante2009}. These
\textit{phenomenological\/} models may be thought of as the simplest
fitting functions available, since they typically include only two
main ingredients: spectral energy distribution (SED) templates, to
relate observed flux densities in different bands given luminosities
and redshifts; and some evolutionary form for the luminosity function,
to produce greater numbers of objects at higher redshifts -- typically
luminosity or density evolution following a power law in ($1+z$).

As observational data at ${\sim}\,10$--1000\,\micron\ have improved,
in terms of wavelength coverage, survey area and depth, many authors
have added greater complexity to their models. For example, it is now
common to divide up the local luminosity function into multiple galaxy
populations, assigning different SEDs to each, and then evolving the
populations independently
\citep[e.g.,][]{rowan-robinson2001,lagache2003}, or assuming some
relation between the IR luminosity and the AGN content
\citep{valiante2009}. With this added freedom, such models can
simultaneously fit the longer-wavelength
850--1200\,\micron\ SCUBA/MAMBO/AzTEC number counts and approximate
redshift distributions, as well as the shallower
\textsl{IRAS/ISO/Spitzer\/} surveys in the FIR (60--200\,\micron), and
more recently the deep 24\,\micron\ \textsl{Spitzer\/} surveys
\citep[e.g.,][]{lagache2004,valiante2009,rowan-robinson2009}.  Despite
these successes, we note that, prior to the first measurements,
predictions for the number counts in the BLAST/SPIRE bands varied
widely \citep[][hereafter \patanchon]{patanchon2009}. We believe there
are two main reasons for these discrepancies. First, the number of
parameters associated with the multiple discrete populations is large,
potentially leading to significant uncertainties in any part of the
spectrum that lacks observational constraints. Second, it is common
knowledge that rest-frame dust temperature, and hence bolometric
luminosity and total dust mass, are degenerate with source redshift
\citep[e.g.,][]{blain2003}; a `redder' object could either be a cooler
local galaxy, or a warmer, more distant galaxy. For this reason,
assumptions about the SED shapes for each galaxy population, and the
potential for evolution in these shapes, can significantly affect the
results.

An alternative to phenomenological modelling of the data is the
\textit{ab initio\/} approach, or solutions to the forward problem:
simulate as much of the physics of galaxy formation and evolution as
possible, and evolve the model forward in time from the Big Bang,
tweaking model parameters to fit observables
\citep[e.g.,][]{baugh2005}. Currently, such models usually incorporate
the numerical evolution of the dark matter distribution, and adopt a
range of recipes to assign galaxies to over-densities in this
distribution, both as a function of mass and the dark matter halo
merger histories. Observational constraints include the cosmic
microwave and infrared backgrounds, the luminosity functions and
spatial clustering of local galaxy populations, and more recently,
their observed surface density and redshift distributions.

For the purpose of understanding galaxy evolution theory, \textit{ab
  initio\/} models presently offer the most complete toolbox, combining
a wide variety of physics into a single coherent model. However, they
are still quite limited in the precision of inference that they allow,
since the simulations lack the necessary resolution, and therefore
assumptions about smaller-scale physics must be made (such as the
details of star formation within molecular clouds, growth of black
holes, and the interaction between the galaxy and the inter-galactic
medium). These assumptions are guided by intuition and simple tuning
recipes based, for example, on information from higher-resolution
hydrodynamic simulations \citep[e.g.,][]{narayanan2010}. These
limitations result in a large number of parameters (undoubtedly with
many partial degeneracies which can be tuned to match the sub-grid
physics), and it can be extremely difficult to estimate meaningful
uncertainties.

In this paper we consider a phenomenological model with more modest
goals. Unlike \textit{ab initio\/} models, and more recent
multi-population phenomenological models that seek to fit the widest
range of data available (e.g., attempting to connect the submm to FIR
and mid-IR galaxy populations, as in
\citealt{chary2001,lagache2004,valiante2009}), we focus our efforts on
data that constrain \textit{only\/} the evolution of the rest-frame
FIR peak ($\lambda \sim 60$--200\,\micron), for galaxies at redshifts
up to $z\sim4.5$. This redshift range encompasses the bulk of the
850--1200\,\micron\ submm galaxy (SMG) population which peaks near
$z\sim2.5$ \citep{aretxaga2003,chapman2003b,chapman2005,chapin2009b},
as well as the most distant spectroscopically-confirmed examples
\citep{capak2008,schinnerer2008,daddi2009,coppin2009}. We are
therefore only attempting to fit data that directly probe
dust-reprocessed radiation from the most active star-forming galaxies,
from their formation epoch to the present day. We also explicitly set
out to determine whether a \emph{single\/} galaxy population with a
simple evolutionary form can reproduce the observed data across the
peak in the CIB. Most authors have concluded, through fitting `by hand',
that multiple populations
with independent evolutionary forms are required by the data;
however, an exhaustive non-linear search of
parameter space has never been performed as in this work to determine:
(i) whether a more complicated model is indeed necessary; and if so
(ii) identify precisely where the tension is coming from to probe the
types of new models, and/or new data that would be required to fit
such models.

For local galaxies, the region of the spectrum we are considering is
quite smooth. At wavelengths $\gsim 100\,\micron$, SED shapes resemble
modified blackbodies, $S_\nu \propto \nu^\beta
B_\nu(T_{\mathrm{obs}})$, although at slightly shorter wavelengths
they are typically brighter and flatter due to a combination of
opacity effects, and sensitivity to warmer dust grains. We therefore
do not need to pay special attention to tuning the MIR spectra of our
models, e.g., stochastically heated small dust grains, including
polycyclic aromatic hydrocarbons, or PAHs \citep{draine2001}, as has
been necessary to fit the deep \textsl{Spitzer\/} 24\,\micron\ data
\citep{chary2001,lagache2004,valiante2009,rowan-robinson2009}. We do,
however, include a range of SED shapes in our analysis, with a
distribution characterised by the single $C \equiv
\log_{10}(S_{60}/S_{100})$ colour near the peak of the rest-frame FIR
emission (which is a good indicator of the FIR peak wavelength) and
its well-known correlation with FIR luminosity, $L \equiv
L_\mathrm{FIR}$, the integrated luminosity from 42.5--122.5\,\micron,
\citep{soifer1991,chapman2003,chapin2009}.

We combine this simple SED shape parameterisation with the local
\textsl{IRAS\/} luminosity function as our local boundary
condition. We then evolve this local bivariate distribution,
$\Phi(L,C)$, using only luminosity and density evolution, to fit the
submm--FIR data. While the idea of incorporating a correlation between
luminosity and FIR colour is not new
\citep[e.g.,][]{lagache2003,lewis2005,valiante2009}, our reduced
number of parameters allows us to fully explore the parameter space
using Monte Carlo Markov Chains (MCMC). To our knowledge, there are
only two other published attempts to fully characterise the
uncertainties in a phenomenological model -- \citet{kelly2008} and
\citet{leborgne2009}, only the latter of which was concerned with
submm--FIR surveys. In addition, a concurrent study by \citet{bethermin2011}
uses methods similar to those presented
here. Another feature of our analysis that sets it apart from earlier
work is our focus on the potential evolution in the correlation
between luminosity and colour, since it is degenerate with redshift
and heavily influences conclusions about the dust-enshrouded
star-formation rate history.

Throughout this paper, we consider two
basic models: one in which the local correlation holds at high
redshift; and a second in which the correlation undergoes luminosity
evolution (making galaxies of a given luminosity in the past appear
cooler than at the present day). This is an area for which BLAST data,
and newer SPIRE data, provide the strongest constraints on this crucial
part of the spectrum. This approach has allowed us both to: choose the
model that best-fits existing data; and clearly indicate what future
data are required to break the remaining degeneracies in parameter
space.

A basic assumption that we make, as with all models of this type, is
that high-redshift luminosity functions smoothly evolve over time to
produce the modern-day $z=0$ luminosity functions. If there is a
significant galaxy population that existed in the early Universe, but
is completely absent in the local Universe (even as a faint tail), our
model will not give plausible results. We also emphasise the fact that
our model \textit{will not\/} give useful predictions for data far
from the rest-frame FIR peak, such as the 24\,\micron\ source
counts. Further work would be needed in order to achieve this, and in
all likelihood, require additional model parameters.

In Sections~\ref{sec:locallc} and \ref{sec:seds} we describe the local
boundary conditions of our model -- the local luminosity and
colour-luminosity distributions, and our adopted SED templates. The
parameterisation of the redshift evolution is described in
Section~\ref{sec:evolmodel}, and the connection of this model to
observed quantities (such as number counts, background intensities and
redshift distributions) is provided in
Section~\ref{sec:observables}. The data sets that we use to fit the
model are given in Section~\ref{sec:data} and our fitting procedure is
summarised in Section~\ref{sec:modelfitting}. The results of the fits
are presented in Section~\ref{sec:results}, and
Section~\ref{sec:discussion} discusses the implications of and future
improvements to the model.

Throughout this paper a standard cosmology is adopted, with
$\Omega_\mathrm{M}=0.272$, $\Omega_\Lambda=0.728$, and
$H_0=70.4$\,km\,s$^{-1}$\,Mpc$^{-1}$ (\citealt{komatsu2011}).

\section{Evolution of the luminosity function}
\label{sec:model}

\subsection{Local Luminosity and Colour Distributions}
\label{sec:locallc}

In most past phenomenological models, authors have used either the
local 60\,\micron\ luminosity function \citep[e.g.,][]{saunders1990}
or the local 850\,\micron\ SCUBA luminosity function
\citep{dunne2000}. The former is one of the most well-studied
luminosity functions, based on the all-sky \textsl{IRAS\/} 12, 25, 60
and 100\,\micron\ survey. Since SCUBA was not sensitive enough to
conduct a survey over a significant portion of the sky, pointed
follow-up of an \textsl{IRAS\/} galaxy sample was employed for the
latter. This technique is adequate, provided that no significant
population of cool ($\la 25$\,K) galaxies exist in the local Universe.
BLAST traces the peak of dust emission at high redshift, as
\textsl{IRAS\/} does locally; emission at 850\,\micron\ comes from the
Rayleigh-Jeans part of the spectrum.  For these reasons, we choose an
\textsl{IRAS\/}-based luminosity function as the basis for our model.

Since we are interested in the range of SED shapes that produce both
the rest-frame submm and FIR emission, we also use the distribution of
$C \equiv \log_{10}(S_{60}/S_{100})$ colours as a function of
luminosity. The observed correlation, with no corrections for
observational biases, was measured by \citet{soifer1991} and has been used in
some phenomenological models \citep[e.g.,][]{lagache2003}. An attempt
was made to measure the full bivariate luminosity-colour distribution
using the $1/V_\mathrm{max}$ technique \citep{schmidt1968} by
\citet{chapman2003}. However, we instead use the updated version of
this distribution from \citet[][hereafter \chapin]{chapin2009}, which
incorporates additional corrections for the \textsl{IRAS\/}
bandpasses, a bias against detecting cooler galaxies in the original
60\,\micron\ flux-limited sample, and redshift evolution. We make one
minor alteration to the distribution; as noted in \chapin\ \citep[and
  other previous authors, e.g.][]{saunders1990,lawrence1999}, the
faint-end of the luminosity function is biased high by the local
over-density of galaxies. Since the joint density of galaxies as a
function of luminosity and colour is formulated in \chapin\ as
$\Phi(L,C) = \Phi(L) \, p(C|L)$ (i.e., the product of a pure
luminosity function, with the conditional probability of a galaxy
having a colour $C$ given a luminosity $L$), we simply replace
$\Phi(L)$ with the measurement from \citet{saunders1990}, which is
valid because they used an estimator that is insensitive to this
over-density (and which was shown in \chapin\ to be consistent at
luminosities $L > L_*$). Here (and throughout), $L$ is defined to be
the integrated 42.5--122.5\,\micron\ FIR luminosity. Throughout, we
follow the convention that $\Phi(L)$ is the number density of objects
per unit luminosity and $\phi(L)$ is the number density per decade of
luminosity.

\subsection{SED Library}
\label{sec:seds}

As mentioned in the introduction, the shapes of the submm--FIR SEDs of
most galaxies in the local Universe are reasonably-well parameterised
by a simple 2-parameter modified blackbody (in addition to the
normalisation), $S_\nu \propto \nu^\beta B(\nu,T)$, where $B(\nu,T)$
is the Planck function. While a range of values of $\beta$ have been
measured, they typically do not vary much from a canonical value of
$\beta=1.5$. Furthermore, $\beta$ and $T_{\mathrm{obs}}$ are highly
anti-correlated in the fits. Therefore, it is plausible that a single
parameter can accurately describe most of the spread in the shapes of
locally-observed FIR SEDs -- in our case we use the FIR colour $C$. In
further support of this simple parameterisation, \citet{dunne2000}
showed that it was possible to map the 60\,\micron\ luminosity
function to the 850\,\micron\ luminosity function by adopting SEDs
that follow the observed correlation between temperature and
luminosity. Similarly, \citet{serjeant2005} fit temperatures to
\textsl{IRAS\/} 60 and 100\,\micron\ data with a fixed $\beta=1.3$,
and for each object estimated their 850\,\micron\ brightnesses. They
found, using these predicted 850\,\micron\ flux densities, that they
could also map the \textsl{IRAS\/} luminosity function to the
850\,\micron\ luminosity function.

In the spirit of these earlier analyses, we seek a set of SED
templates that can transform the \textsl{IRAS\/} colour-luminosity
distribution to local luminosity functions at other adjacent
wavelengths, namely: \textsl{IRAS\/} 12\,\micron\ \citep{fang1998};
ISOCAM 15\,\micron\ \citep{xu2000}; \textsl{IRAS\/}
25\,\micron\ \citep{shupe1998}; and SCUBA
850\,\micron\ \citep{dunne2001}.\footnote{Since each study adopted
  slightly different values of $H_0$, we have corrected the
  luminosities and volumes for the value
  $H_0=70.4$\,km\,s$^{-1}$\,Mpc$^{-1}$ used in this paper.} The
shortest wavelength data that we will attempt to fit are the observed
source counts at 70\,\micron. It is therefore important to match the
local luminosity functions at wavelengths as short as 12\,\micron,
since this wavelength is redshifted to 70\,\micron\ at $z\,{\sim}\,5$
(we only expect a minor contribution to the 12--15\,\micron\ SEDs from
more complicated emission mechanisms, and this will only impact the
70\,\micron\ measurements for $z \gsim 4.5$ sources). On the
long-wavelength side, by achieving consistency between the
\textsl{IRAS\/} and 850\,\micron\ luminosity functions, we can expect
to reasonably interpolate the rest-frame luminosity functions in the
BLAST bands that are bracketed by these wavelengths. Again, we
emphasise that we are \textit{not\/} attempting to fit
24\,\micron\ number counts, which would require accurate modelling of
the data at shorter wavelengths ($\lsim8$\,\micron) where the SEDs are
considerably more complicated.

We have examined and rejected three common SED models for individual
galaxies. First, we attempted to use single-temperature modified
blackbodies. We produced a library of SEDs with fixed values of
$\beta$ and a range of temperatures. We then predicted each of the
monochromatic luminosity functions, $\Phi(L_\nu)$, from $\Phi(L,C)$,
\begin{equation}
\label{eqn:lfuncmodel}
\Phi(L_\nu) = \int \Phi( L(L_\nu,C), C ) \, \frac{\partial L(L_\nu,C)}{\partial L_\nu} \, dC,
\end{equation}
where $L(L_\nu,C)$ is the FIR luminosity for an SED in our library
of colour $C$, normalised to the luminosity density $L_\nu$ at a
frequency $\nu$. Similar to \citet{dunne2000} and
\citet{serjeant2005}, we were able to obtain good agreement between
the \textsl{IRAS\/} and 850\,\micron\ luminosity functions using
$\beta=1.5$. However, unsurprisingly, this simplistic model is a poor
fit to the shorter wavelength data, since the Wien tail of our
single-temperature SEDs falls off considerably more rapidly than for
real galaxies, which contain a mixture of dust at different
temperatures and compositions. Even the apparent plausibility of our
SEDs for the longer-wavelength data can be deceiving; an effective
$\beta=1.5$ single-temperature spectrum can also be produced by the
superposition of a range of dust populations with a steeper
$\beta=2.0$ at a range of appropriately-selected temperatures. This
fact serves to remind us that no simple physical meaning should be
attached to these model parameters; the modified blackbody is only a
convenient fitting function.

Next, we tested two more realistic SED libraries (each spanning the
submm--IR wavelengths of interest) that are commonly used in the submm
and FIR literature: the templates from \citet{chary2001} that were fit
to data spanning 0.44--850\,\micron, and which were used in their
phenomenological model, based on evolution of the
15\,\micron\ luminosity function; and the templates of star-forming
galaxies from \citet{dale2001} that were fit to \textsl{IRAS\/} and
\textsl{ISO\/} data. The \citet{chary2001} templates provided a
reasonable extrapolation to the luminosity functions at
12--25\,\micron, but led to a significant over-prediction at
850\,\micron. The \citet{dale2001} SEDs performed much better at
850\,\micron, but led to moderate over-predictions at
12--25\,\micron. Due to these shortcomings, we decided to produce our
own SED templates that vary smoothly as a function of $C$.

The basis of our SED library is the model of \citet{draine2007}. Their
parameterisation includes: (i) a set of mid-IR templates as a function
of the PAH abundance, $q_\mathrm{PAH}$; and (ii) longer-wavelength
templates for cooler thermal emission that is composed of dust heated
both by a single low-intensity radiation field, $U_\mathrm{min}$, and
a second component heated by a range of radiation intensities from
$U_\mathrm{min}$ to $U_\mathrm{max}$, where a factor $\gamma$ gives
the fraction of the total dust emission produced by this second
component. We experimented with these 4 parameters ($q_\mathrm{PAH}$,
$U_\mathrm{min}$, $U_\mathrm{max}$ and $\gamma$) to produce a sequence
of smoothly varying SEDs as a function of $C$ that resulted in good
extrapolations to the monochromatic luminosity functions on both sides
of the FIR peak.

\begin{figure}
\centering
\includegraphics[width=\linewidth]{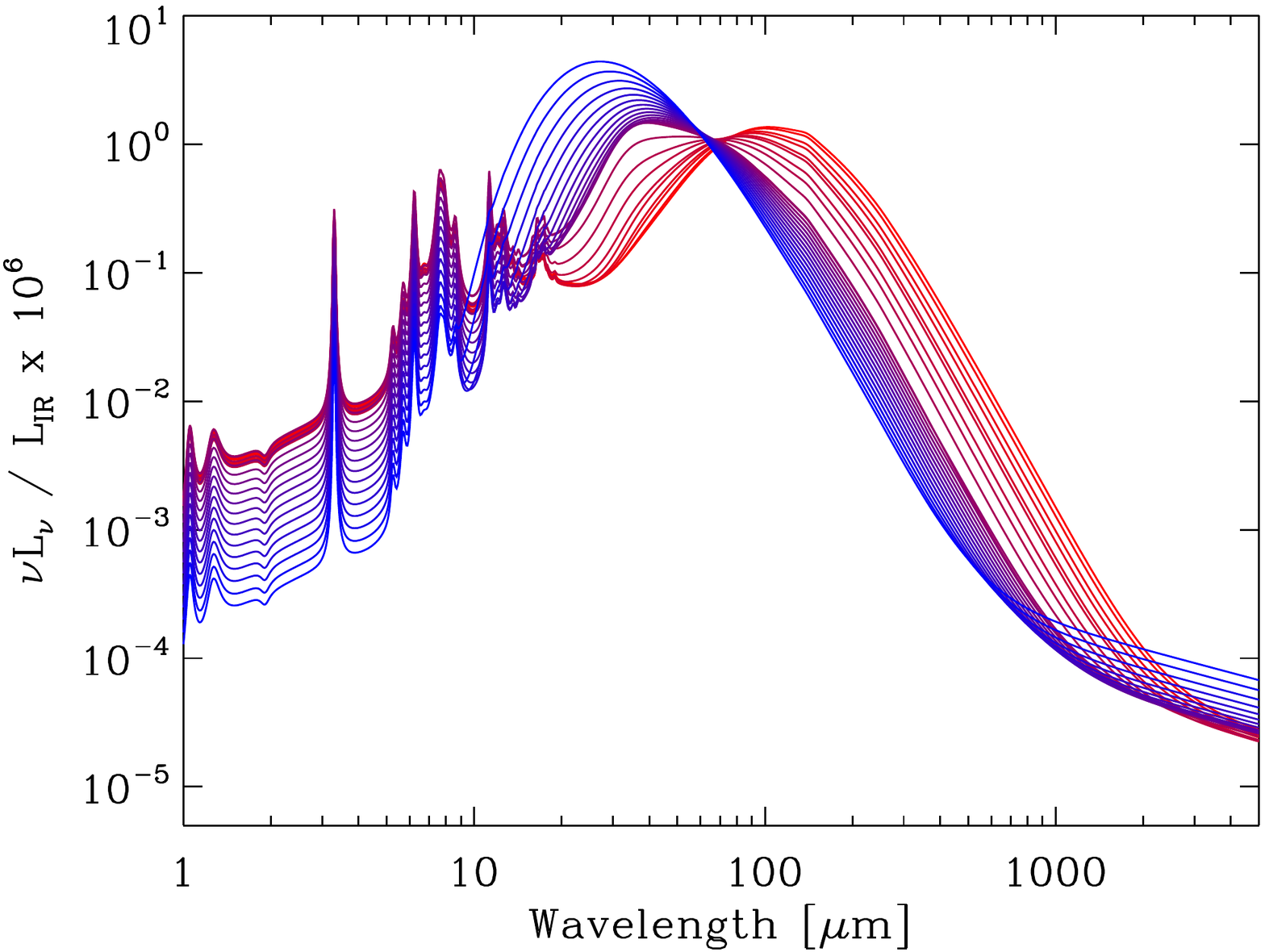}
\caption{SEDs used in our model, generated from a range of templates
  presented in \citet{draine2007}. The 60-to-100\,\micron\ colour $C$
  of these SEDs ranges from $-0.5$ to 0.5, indicated by a colour
  gradient from red (cool, $C=-0.5$) to blue (warm, $C=0.5$). The
  abrupt increase in the density of SEDs which peak at $\lambda \lsim
  30$\,\micron\ is caused by an extrapolation from the warmest
  \citet{draine2007} model (with $C=0.1$) to even larger values of $C$
  by adding modified blackbodies with $\beta=1.5$ and temperatures
  ranging from 47--98\,K. A radio component is added on, based on the
  FIR-radio correlations; this dominates at $\lambda \gsim
  0.5$--2\,mm.}
\label{fig:seds}
\end{figure}

We found that the results did not depend particularly heavily on
$q_\mathrm{PAH}$ and so thereafter we simply fix it to an intermediate
value of 2.50 (from a possible range spanning 0.10--4.58). The value
of $U_\mathrm{min}$ effectively sets the apparent temperature of the
coolest dust, and larger values of $U_\mathrm{max}$ and $\gamma$
increase the temperature and fraction of the hotter dust (i.e.,
together these parameters control most of the submm--FIR SED
shape). We obtained good fits for our extrapolated luminosity
functions by fixing $U_\mathrm{max}=10^{4}$ (from a possible range
spanning $10^3$--$10^6$), and stepping through all 22 of the supplied
templates, corresponding to values of $0.1 \le U_\mathrm{min} \le
25.0$. We simultaneously increased $\gamma$ logarithmically from
$10^{-9}$ to 0.4 over the 22 levels. Finally, for this set of SEDs, we
found that the values of $C$ only spanned $-0.50$ to 0.10, thereby
missing some of the warmest values identified in the local Universe,
and to a lesser extent, some of the cooler values (see fig.~4 in
\chapin). Since the Rayleigh-Jeans side of the \citet{draine2007}
models resemble modified blackbodies with $\beta=1.5$, we simply
extended the library to a larger value of $C=0.5$ by taking the
$C=0.10$ SED and adding on modified blackbodies with $\beta=1.5$,
temperatures $T$ ranging from 47--98\,K, and normalised such that they
pass through the $C=0.10$ model at 100\,\micron. At the red end, we
simply assign the $C=-0.50$ SED to galaxies with $C < -0.50$ (i.e.,
galaxies that appear to have temperatures $T \lsim 27$\,K using a
$\beta=1.5$ modified blackbody are truncated at $T =
27$\,K). Fig.~\ref{fig:seds} shows the complete set of SEDs. The
transition to the extrapolated $C>0.10$ SEDs are obvious as an abrupt
increase in the density of templates that peak at wavelengths $\lambda
\lsim 30$\,\micron.

While it does not affect the results of the model fits, we also add on
a power-law radio component to the SEDs, based on the FIR-radio
correlation of \citet{ivison2010}. These authors measure the quantity
$q_\mathrm{IR}$, the ratio of the rest-frame 8--1000\,\micron\ flux to
the 1400\,MHz flux density,
\begin{equation}
q_\mathrm{IR} \equiv \log \left(\frac{S_\mathrm{IR} / 3.75 \times 10^{12}}{\mathrm{W\,m^{-2}}} \right) - \log \left(\frac{S_\mathrm{1400\,MHz}}{\mathrm{W\,m^{-2}\,Hz^{-1}}} \right),
\end{equation}
and also a power law index $\alpha_\mathrm{rad}$. We add the radio
continuum using $q_\mathrm{IR}=2.41$ and $\alpha_\mathrm{rad} = -0.4$
at wavelengths longer than 100\,\micron.\footnote{The quantity
  $q_\mathrm{IR}$ has been updated to 2.40 using
  \textsl{Herschel\/}/SPIRE observations \citep{ivison2010b}, but as
  the change is significantly smaller than the measurement errors, we
  have not updated our value.} This has a small effect on the number
counts predictions at millimetre wavelengths, presented in
Section~\ref{sec:results}.

\begin{figure}
\centering
\includegraphics[width=\linewidth]{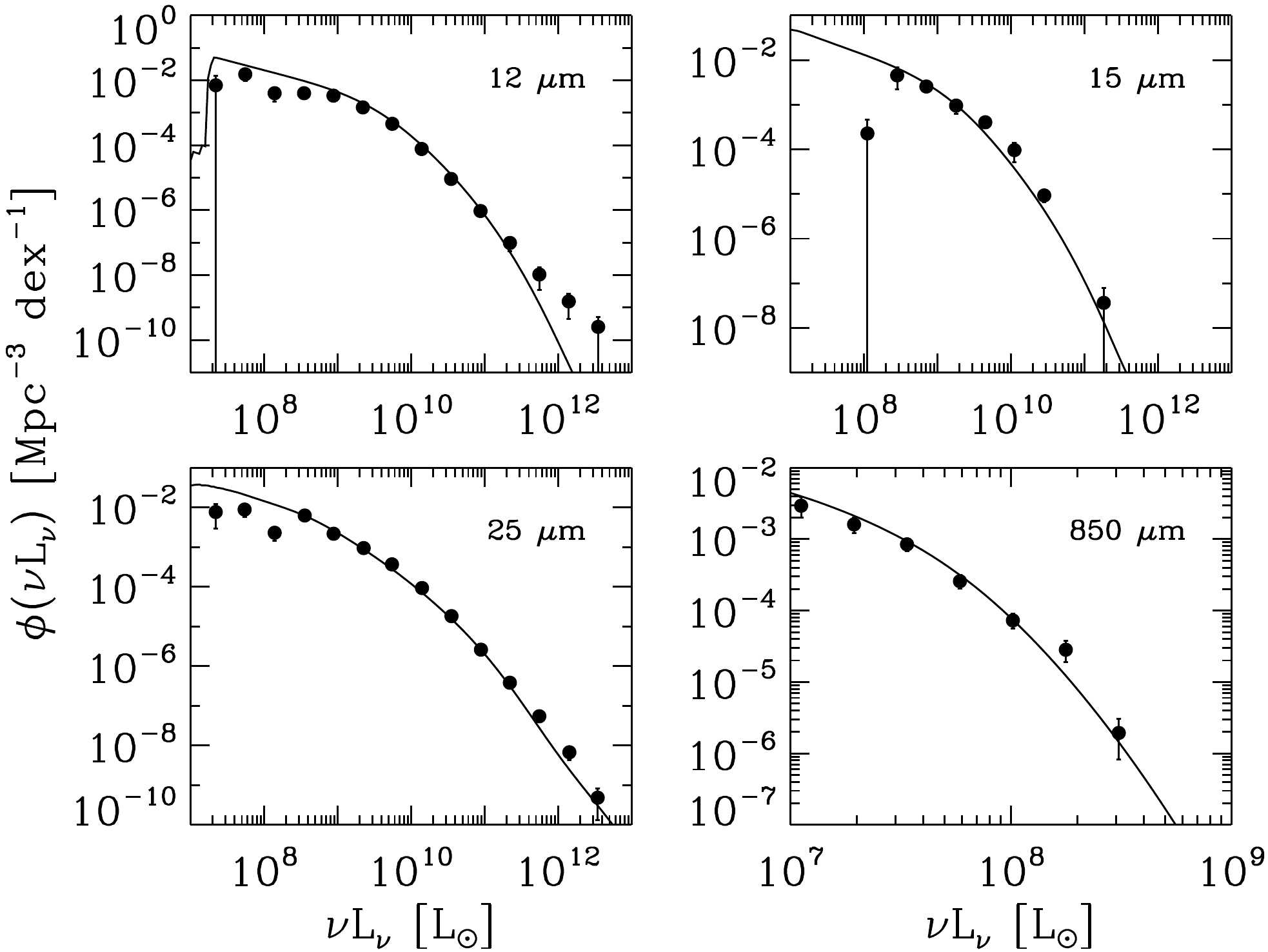}
\caption{The local luminosity density functions derived from our
  adopted local luminosity function $\Phi(L)$, colour-luminosity
  function $p(C|L)$ and SED library, described in
  Sections~\ref{sec:locallc} and \ref{sec:seds}. The 12, 15, 25 and
  850\,\micron\ data sets are from \citet{fang1998}, \citet{xu2000},
  \citet{shupe1998} and \citet{dunne2000}, respectively.}
\label{fig:locallumfunc}
\end{figure}

Fig.~\ref{fig:locallumfunc} shows the derived local monochromatic
luminosity functions using the above SED library. The good
correspondence indicates that we have found a plausible 2-parameter
model for the distribution of galaxy luminosities and colours spanning
12--850\,\micron. Later, in Section~\ref{sec:colcol}, we offer further
evidence that this SED library is reasonable over the wavelength range
of interest by comparing the colours of galaxies sampled from our
best-fit model with real surveys that span 24--850\,\micron.

\subsection{Extension to High Redshift}
\label{sec:evolmodel}

We use a simple extension of the local luminosity function $\Phi(L,C)$
to high redshift, incorporating parametric forms for density
$\rho(z)$, luminosity $g(z)$ and colour-luminosity $h(z)$ evolution:
\begin{equation}
\label{eqn:evollumfunc}
\Phi(L,C,z) = \rho(z)\times\Phi\left(\frac{L}{g(z)}\right)\times
p\left(C\left|\frac{L}{h(z)}\right.\right).
\end{equation}
Here, $\Phi(L,C,z)$ is the comoving luminosity evolution function,
with units L$_\odot^{-1}$\,Mpc$^{-3}$. Then the number of galaxies in a
cell centred at ($L,C,z)$ is:
\begin{equation}
 N = \Phi(L,C,z)\,\Delta L\,\Delta C\,\frac{dV}{dz}\,\Delta z,
\end{equation}
where $\Delta L$, $\Delta C$ and $\Delta z$ are the dimensions of the cell.

We note that if $\Phi(L)$ were a power law, $\rho$ and $g$ would be
completely degenerate \citep[this is a well-known problem,
  e.g.,][]{saunders1990}. Our $\Phi(L)$ has both a break (at $L_* = 4
\times 10^9\,\mathrm{L}_\odot$) and a low-luminosity cut-off (at
$L_\mathrm{cut} = 10^8\,\mathrm{L}_\odot$). These features, combined
with our use of a colour-luminosity correlation, serve to break some
of the degeneracy between the parametric functions, although strong
correlations remain. We have not fully tested the effect of varying the
low-luminosity cut-off (although see Section~\ref{sec:reconciling}),
but we note that the total luminosity, $\int L
\Phi(L) dL$, integrated from $L_\mathrm{cut}$ to infinity, is
${\sim}\,90$ per cent of the full integral.

To reduce the number of free parameters, we set $h(z) =
[g(z)]^\alpha$. The parameter $\alpha$ controls the amount of
colour-luminosity evolution in the model: with $\alpha = 1$, the
right-hand side of Equation~\ref{eqn:evollumfunc} can be written as
$\rho(z) \times\Phi(L/g(z),C)$; with $\alpha = 0$, the
colour-luminosity relation does not evolve with redshift. The
parameterisation of $\rho(z)$ and $g(z)$ is discussed in
Section~\ref{sec:parametrization}. We explore models with $\alpha$
fixed to 1.0 and 0.0, as well as with $\alpha$ allowed to vary as a
free parameter. In all cases, $\rho$ and $g$ are free parameterised
functions. The consequences of varying $\alpha$ are significant and
are discussed in Section~\ref{sec:clevol}.

\subsection{Observables}
\label{sec:observables}

Our evolving luminosity function can be integrated across the
appropriate variables to provide observables from the model. We first
change variables from intrinsic luminosity $L$ to observed flux
density $S_\nu$:
\begin{equation}
f(S_\nu,C,z) = 
\Phi \left(L(S_\nu,C,z),C,z\right)
\left(\frac{\partial L}{\partial S_\nu}\right)\left(\frac{dV}{dz}\right),
\end{equation}
with
\begin{equation}
L(S_\nu,C,z) = \frac{4 \pi D_\mathrm{L}^2(z) \,S_\nu}{(1+z)\,T(C,(1+z) \nu)},
\end{equation}
where $D_\mathrm{L}$ is the luminosity distance and $T(C,\nu)$
converts FIR luminosity $L$ to luminosity density $L_\nu$ at
rest-frame frequency $\nu$ for the SED template with colour $C$. The
$dV/dz$ term converts the counts from number per unit volume to number
per unit redshift.

We use four types of data in this analysis:

\begin{itemize}

\item differential number counts, calculated by integrating across
  colour and redshift.
\begin{equation}
\label{eqn:counts}
\frac{dN(S_\nu)}{dS_\nu} = \int_0^\infty \int_{-\infty}^\infty
f(S_\nu,C,z) \,dC\,dz;
\end{equation}

\item background intensity (CIB), obtained by further integration over $S_\nu$,
\begin{equation}
\label{eqn:cib}
I_\nu = \int_0^\infty S_\nu
\left[\frac{dN(S_\nu)}{dS_\nu}\right]\, dS_\nu;
\end{equation}

\item background intensity as a function of redshift of emitting sources,
\begin{equation}
\label{eqn:diffintensity}
\frac{dI_\nu}{dz} = \int_0^\infty \int_{-\infty}^\infty
S_\nu \,f(S_\nu,C,z) \,dS_\nu\,dC;
\end{equation}

\item number of sources brighter than $S_\mathrm{lim}$ as a function of redshift,
\begin{equation}
\label{eqn:zcounts}
\left. \frac{dN}{dz}\right|_{S_\nu > S_\mathrm{lim}} = \int_{S_\mathrm{lim}}^\infty \int_{-\infty}^\infty 
f(S_\nu,C,z) \,dS_\nu\,dC.
\end{equation}

\end{itemize}

\section{Data}
\label{sec:data}

We now describe each of the data sets which we use to constrain the model.

\subsection{Number Counts}

From the choice of possible number counts, we use the following.

\begin{enumerate}
\item \textsl{Spitzer\/} MIPS counts at 70 and
  160\,\micron\ \citep{bethermin2010}.

\item BLAST counts from table~3 of \patanchon\ (not constrained by the
  CIB) with covariance matrices (upper quadrants of tables~4--6 in
  \patanchon). Since the BLAST counts are given as a series of nodes
  connected by power laws and not counts-in-bins, we treat these data
  differently than counts from other instruments. This is discussed
  further in Section~\ref{sec:likelicalc}. We have not accounted for
  the ${\sim}\,10$ per cent calibration errors, which are strongly
  correlated \citep{truch2009}, or for errors due to cosmic variance,
  which may be significant at the bright end.

\item AzTEC 1.1\,mm counts in the SHADES fields presented in
  \citet{austermann2010}. A covariance matrix is given; however, the
  correlations are very high, and the paper warns about
  over-interpreting measured correlations, so we use only the diagonal
  elements of the covariance matrix, ignoring correlations. This may
  over-weight the AzTEC counts, although a simple test (fitting the
  amplitude of a Schechter function with fixed shape) shows that the
  errors are reasonable.
\end{enumerate}

\begin{figure}
\centering
\includegraphics[width=\linewidth]{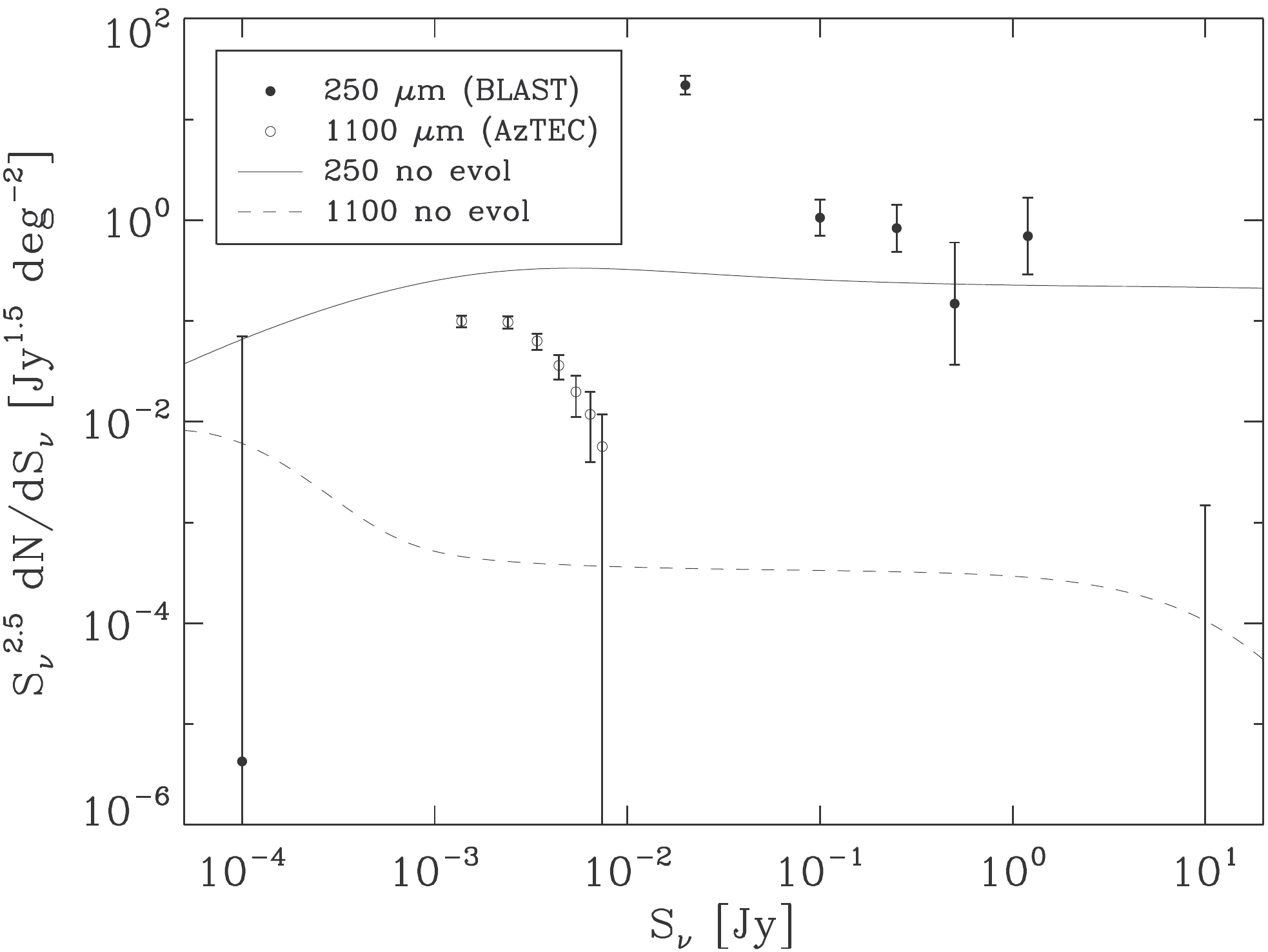}
\caption{No-evolution models compared to data at 250 and
  1100\,\micron. The error bars at $10^{-4}$ and $10^1$ are part of
  the BLAST 250\,\micron\ data set.}
\label{fig:counts_noevol}
\end{figure}

In Fig.~\ref{fig:counts_noevol}, we show the no-evolution counts
(Equation~\ref{eqn:evollumfunc} with the evolutionary parameters
$\rho$, $g$ and $h$ all set to 1.0). These are derived from the
2-parameter local luminosity function described in
Section~\ref{sec:locallc} at 250 and 1100\,\micron\ compared to the
measured counts by BLAST and AzTEC, respectively. We see that while
the 250\,\micron\ counts are Euclidean at the bright end, they show
strong evolution in the $S_\nu=10$--100\,mJy range. The
1100\,\micron\ counts show strong evolution over their full range.

\subsection{Background Intensity}

We use observations of the CIB reported by \citet{fixsen1998}. We
choose to fit the model at the same wavelengths at which the counts
are used: 160, 250, 350, 500, 850 and 1100\,\micron. We assume 30 per
cent band-independent errors (discussed further in
Section~\ref{sec:goodnessoffit}).

\subsection{Redshift Distribution}
\label{sec:redshiftdist}

We use $dN/dz$ for SCUBA galaxies as measured by
\citet{chapman2005}. They present a histogram of 73 sources (their
fig.~4), but over-plot a model to show `the likely effects of the
sample selection.' We assign each of the 9 histogram bins a Poisson
error, then scale the histogram bin values and errors to fit their
model. We fit a 3-parameter Gaussian to the histogram and use the
fitted amplitude to normalise the histogram. These scaled bin values
and errors are used to constrain the shape of the model redshift
distribution. We assume a limiting flux density $S_\mathrm{lim}$ of 5\,mJy.

\subsection{Data Sets Not Used}
\label{sec:datanotused}

A number of other relevant data sets exist that, for various reasons,
we do not use to constrain the model. In most cases, the predictions
of the best-fit model are compared to the unused data sets in
Section~\ref{sec:results}.

\subsubsection{Counts}
\label{sec:omitcounts}

We have chosen to omit the SCUBA 850\,\micron\ number counts
\citep[e.g.,][]{coppin2006} due to the fact that there is considerable
tension between these measurements and those performed more recently
with the AzTEC camera in a number of different fields at 1.1\,mm. It
has been noted since the very first surveys were undertaken with AzTEC
that in order to scale its observed galaxy counts to those observed
with SCUBA, each galaxy, roughly speaking, would have to be a factor
of $\sim$3 brighter at 850\,\micron\ \citep[e.g.][]{perera2008},
whereas on an object-by-object basis, the individual galaxies appear
to have a flux density ratio closer to 2, which is about what would be
expected for a typical SED at redshift $\sim$2
\citep[e.g.][]{perera2008,chapin2009}. By itself, this comparison
hints at a bias in counting, even though individual objects appear to
be well-behaved. We note that as counting methods have improved, the
downward correction for biases in the SCUBA counts has increased; the
\citet{coppin2006} measurement falls below essentially all of the
previous estimates at $S_{850} > 2$\,mJy, which were made using simpler
techniques. Similarly, the methodology followed by the AzTEC team has
also evolved over time and more biases have been discovered and
corrected tending to lower the counts
\citep[e.g.][]{austermann2010,downes2011}. We also note that the
counts at 870\,\micron\ measured with LABOCA fall considerably below
those measured with SCUBA \citep{weiss2009}.

We have found that it is impossible to fit a model that is consistent
with the AzTEC 1.1\,mm and SCUBA 850\,\micron\ counts simultaneously
within the quoted uncertainties. We note that most recent
phenomenological models have also only attempted to fit SCUBA counts,
rather than including other counts near 1\,mm
\citep[e.g.][]{lagache2003,lagache2004,valiante2009,leborgne2009,bethermin2011}.
We have explicitly checked that the model of \citet{valiante2009}, which
fits the SCUBA 850\,\micron\ counts, significantly exceeds the AzTEC
1.1\,mm counts, as we find here.

Taking all of these facts into consideration, there is strong evidence
that the counts near 1\,mm are biased by amounts that are not
accurately characterised by published uncertainties. We feel that the
best results will be obtained using the more recent, and more
sophisticated, AzTEC measurements. This, however, is clearly an open
subject that needs to be fully addressed in future assessments.

Number counts at 1.4 and 2.0\,mm measured by the South Pole Telescope
(SPT) were recently published by \citet{vieira2010}. They measured the
bright end of the number counts, however, for which contributions by
lensed galaxies are expected. Since our model does not include
lensing, we do not use these counts to constrain our model.

\subsubsection{Intensity in Redshift Slices} 

\citet[][table~2]{pascale2009} presents the CIB at the BLAST
wavelengths due to 24\,\micron\ sources in six redshift bins from
$z\,{\sim}\,0.4$--2. We attempted to use this data set to constrain
the evolution model by comparing the values to the integral of the
model over each redshift bin, but had difficulty reconciling this data
set with the others; we found that, particularly at $z<0.5$, the model
preferred significantly lower values than the BLAST
measurements. Since the analysis uses photometric redshifts, which may
be unreliable, we have not included these data in our fits. We also
note that the numbers of galaxies in their sample at these lower
redshifts are extremely small; sampling variance due to clustering
is certainly a large
term that must be added to the quoted Poisson uncertainties.

\subsubsection{Redshift Distribution}
\label{sec:reddist}

A number of measurements of redshifts distributions have been made,
including at: 170\,\micron\ by \textsl{ISO\/}
\citep{patris2003,dennefeld2005,taylor2005}; 250, 350 and
500\,\micron\ by BLAST \citep{dye2009,dunlop2010,chapin2011}; and
1.1\,mm by AzTEC \citep{chapin2009b}. We assume limiting flux densities
$S_\mathrm{lim}$ of 200, 40, 20 and 3.8\,mJy for the 170, 250, 500
and 1100\,\micron\ distributions, respectively (we have not considered
the 350\,\micron\ data set). We do not fit these $dN/dz$
distributions, as the selection biases in producing the catalogues,
due to the need for an optical counterpart to identify each source
redshift, are difficult to quantify. This is also the case for the
SCUBA distribution that we \textit{do\/} use (although an attempt to
correct the bias has been applied); however, we wanted to include at
least one data set to constrain the redshift distribution, since the
model is degenerate without it, and the SCUBA sample is the largest
available.

\section{Model fitting}
\label{sec:modelfitting}

Given the data sets listed in Section~\ref{sec:data}, we map out the
likelihood space of the luminosity evolution model using MCMC. This
allows us to quote most-likely parameter values along with errors and
correlations between parameters.

\subsection{Parameterisation}
\label{sec:parametrization}

We parameterise the evolution functions $\rho(z)$ and $g(z)$ as
connected power laws at a series of nodes at a specific set of
$z_i$. We have chosen 6 free parameters in each of $\rho$ and $z$ at a
set of $z_i$ spaced roughly linearly in $\log(1+z)$, spanning
$z=0$--5; the values used are $z_i=$0.1, 0.5, 1.0, 2.0, 3.5 and
5.0. The functions $\rho$ and $g$ are both fixed to 1.0 at $z=0$ and
to $10^{-12}$ at $z=7$, which serves as a high-redshift cutoff. The
evolution parameters $\log(\rho_i)$ and $\log(g_i)$ are then linearly
interpolated in $\log(1+z)$.

\subsection{Likelihood Calculations}
\label{sec:likelicalc}

We calculate the likelihood of a model with a given set of evolution
parameters based on $\chi^2$,
\begin{equation}
{\cal L}(\{\xi_j\}) = \exp \left(- \frac{1}{2} \sum_i w_i
\chi_\mathrm{i}^2 \right),
\end{equation}
where $\{\xi_j\}$ represents the $N_\mathrm{p}$ free parameters and
the sum is over the $N_\mathrm{d}$ data sets. This formulation
inherently assumes that the errors on the data sets are Gaussian
distributed; this is not always a valid assumption, but it allows us
to proceed in a straightforward manner.

In general, counts are treated as counts-in-bins. At each bin centre,
we compare the model (Equation~\ref{eqn:counts}) to the data. The full
covariance matrix is used if available. However, as previously
discussed, the BLAST $P(D)$ counts are treated differently: both the
model and the data (connected power laws) are integrated between each
pair of nodes, and the integrated values are compared. Errors
(including correlations, which can be significant) on the
data integrals are measured using Monte Carlo simulations, sampling
from the chains produced by \patanchon. We use Gaussians centred on
the median values and with widths equal to half of the 68 per cent
confidence regions. This is a reasonable description for all values
except for the integrals between the two faintest bins at each
wavelength, which have large positive tails.

For the SCUBA $dN/dz$, we compare the model to the normalised Gaussian
discussed in Section~\ref{sec:redshiftdist} at the six redshift points
$z_i$ defined above. Errors at these points are obtained by
propagating errors, including correlations, measured by the MCMC.

We are, in principle, free to set the relative weights $w_i$ of each
data set, but for now the weights are ignored ($w_i = 1$) and the
relative importance of each data set depends on the number of
measurements (and their errors) in each set.

\subsection{Monte Carlos}

We use CosmoMC\footnote{\url{http://cosmologist.info/cosmomc/}} as our
likelihood sampler \citep{lewis2002}, using Metropolis-Hastings
sampling. We run each model using 4 chains and run to an `R-1'
convergence of 0.003 to ensure accurate confidence limits. This
typically requires ${\sim}\,10^5$ samples per chain.

\section{Results}
\label{sec:results}

We fit 12 evolutionary parameters ($\rho_i$ and $g_i$ at the six $z_i$)
to the data sets listed in Section~\ref{sec:data}. We find an
important dependence on the parameter $\alpha$, which governs the
extent of evolution in the colour-luminosity relationship
(Section~\ref{sec:evolmodel}); we have tested both $\alpha = 1$
(colour-luminosity evolution) and $\alpha = 0$ (no colour-luminosity
evolution) and find that the $\alpha = 1$ model is a better fit to
the data. We therefore concentrate on the $\alpha = 1$ model, but also
show results for the $\alpha = 0$ model for comparison. The results of
the MCMC analysis are presented in
Figs.~\ref{fig:params}--\ref{fig:nzdist}. Throughout, the results for
the $\alpha=1$ model are in blue and for $\alpha = 0$ in red, with the
best-fit models in thick solid and thick dashed lines,
respectively. Realizations of the model, including luminosity
functions, counts and sample sources lists, are available at
\url{http://cmbr.phas.ubc.ca/model/}.

The implications of the choice for $\alpha$, along with fits with
$\alpha$ as a free parameter, are discussed further in
Section~\ref{sec:clevol}. Although it is it not shown here, the free
$\alpha$ model leads to an intermediate value of $\alpha = 0.62 \pm
0.04$, with correlation coefficients (to the other parameters) as
large as 0.6; the correlations are largest for the low-redshift
parameters and are nearly zero for the high-redshift parameters ($z_i
> 1$).

\subsection{Parameters}

\begin{figure}
\centering
\includegraphics[width=\linewidth]{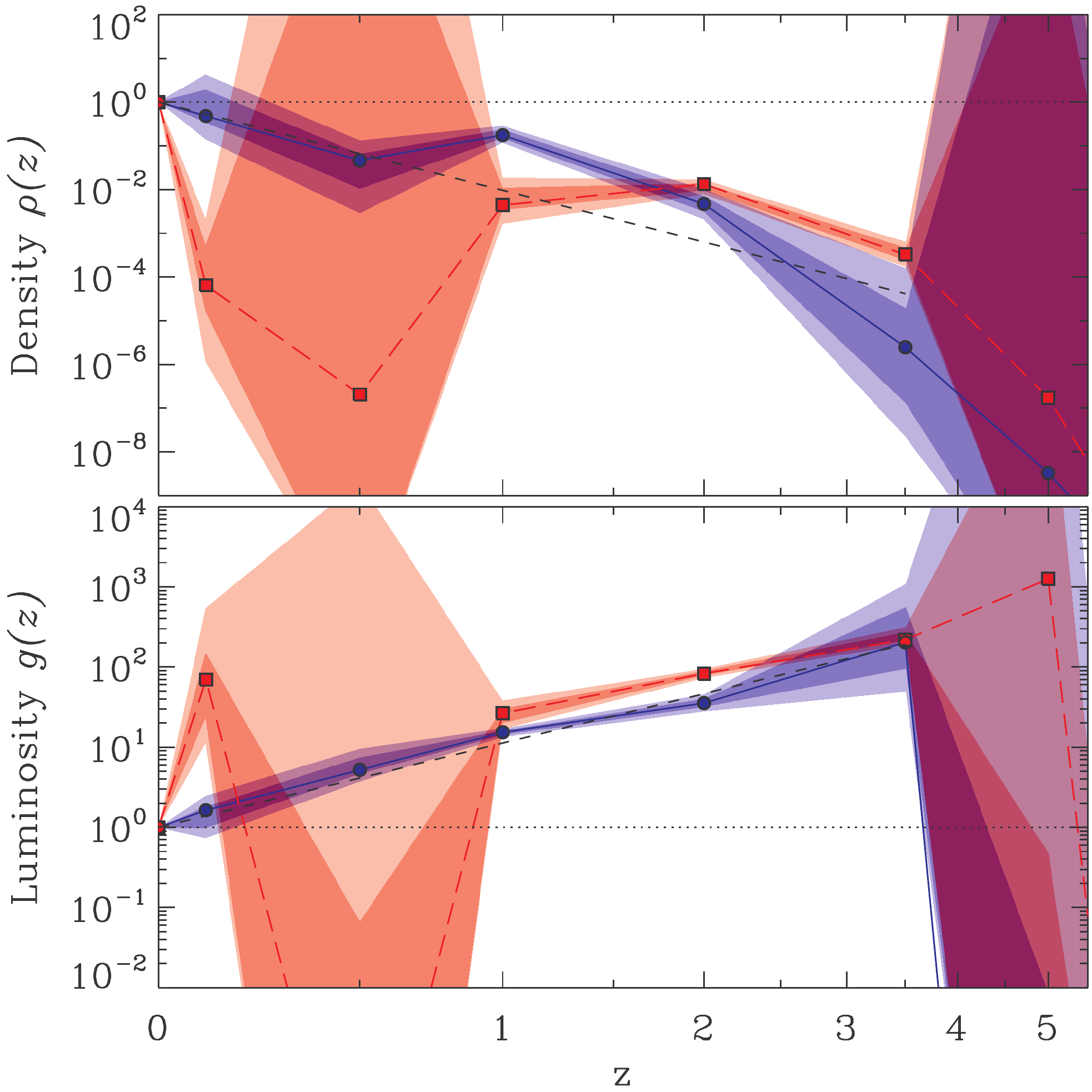}
\caption{The parameter values, which are fit to the data sets shown in
  Figs.~\ref{fig:counts1}--\ref{fig:nzdist}. Above, density evolution,
  $\rho(z)$, and below, luminosity evolution $g(z)$. The $C$--$L$
  evolution ($\alpha = 1$) and no $C$--$L$ evolution ($\alpha = 0$)
  models are shown, in blue and red, respectively. The values of
  $\rho$ and $g$ at $z=0$ and $z=7$ are fixed to 1.0 and 10$^{-12}$,
  respectively. The values in between, at $z$=0.1, 0.5, 1.0, 2.0, 3.5
  and 5.0, are free parameters in the model. The blue (red) symbols
  connected by solid (long-dashed) lines indicate the best-fit
  parameter values for the $\alpha=1$ ($\alpha=0$) models. The
  coloured bands represent the 68 and 95 per cent confidence
  regions. We note that $\rho_i$ and $g_i$ at each $z_i$ are highly
  anti-correlated (see Fig.~\ref{fig:chains}). Representative
  power laws, $\rho(z) = (1+z)^\gamma$ and $g(z) = (1+z)^\delta$ up to
  a cutoff redshift $z=3.5$, with $\gamma = -6.7$ and $\delta = 3.5$,
  have been over-plotted as black short-dashed lines.}
\label{fig:params}
\end{figure}

The best-fit evolution functions $\rho(z)$ and $g(z)$, along with 68
and 95 per cent confidence intervals, are shown in
Fig.~\ref{fig:params}. Both models show clear negative density
evolution and positive luminosity evolution with increasing
redshift. Both models remain well constrained up to $z=3.5$, with the
exception of the $z=0.5$ node in the $\alpha = 0$ model; this feature
is discussed further in Section~\ref{sec:rescounts}. The data do not
constrain the models above $z=3.5$.

To guide the eye, we show (as a thin-dashed line) for both density and
luminosity functions a representative single power law, $\rho(z) =
(1+z)^\gamma$ and $g(z) = (1+z)^\delta$ up to a cutoff redshift
$z=3.5$. These values have not been fit to the data, but instead fit
to the best-fit $\alpha=1$ values. The power indices of the lines
shown are $\gamma = -6.7$ and $\delta = 3.5$. A generic feature of our
models is therefore a combined trend of \textit{negative density
  evolution\/} and \textit{positive luminosity evolution\/} as one
observes FIR-selected galaxies further into the past.

Density evolution may be considered an indication of the overall
galaxy merger rate as a function of time. If galaxy formation were to
follow a simple bottom-up scenario, such as in the case of dark matter
halo merger histories, the smallest bodies are created first, and over
time merge together to form a smaller number of galaxies (i.e.,
positive density evolution). Luminosity evolution does not change the
total number of galaxies in the Universe, but rather their brightness
distribution. Under the previous scenario, one might naively expect a
large number of less-luminous galaxies to merge together in the past,
and produce a smaller number of more-luminous sources in the present
(i.e., negative luminosity evolution combined with positive density
evolution). However, precisely the opposite behaviour is observed in
deep extra-galactic surveys at different wavelengths. This apparent
`cosmic downsizing' \citep[as first noted by][]{cowie1996} has been a
topic of great interest to theorists, and our modelling results
continue to support the trends observed at other wavelengths:
regardless of the parameterisation, there were fewer, but more
luminous FIR-bright galaxies in the past. For the most luminous of
these galaxies to have formed in the early Universe, most of their
stars must have been created in a relatively short burst (less than
1\,Gyr for galaxies at $z\gsim3.0$, given the available time since the
first generation of stars during re-ionisation). The FIR-luminous
galaxies presumably faded slowly over time as the more massive,
luminous stars completed their life-cycle, giving way to lower-mass,
cooler and less-luminous stars.

\begin{figure*} 
\centering
\includegraphics[width=\linewidth]{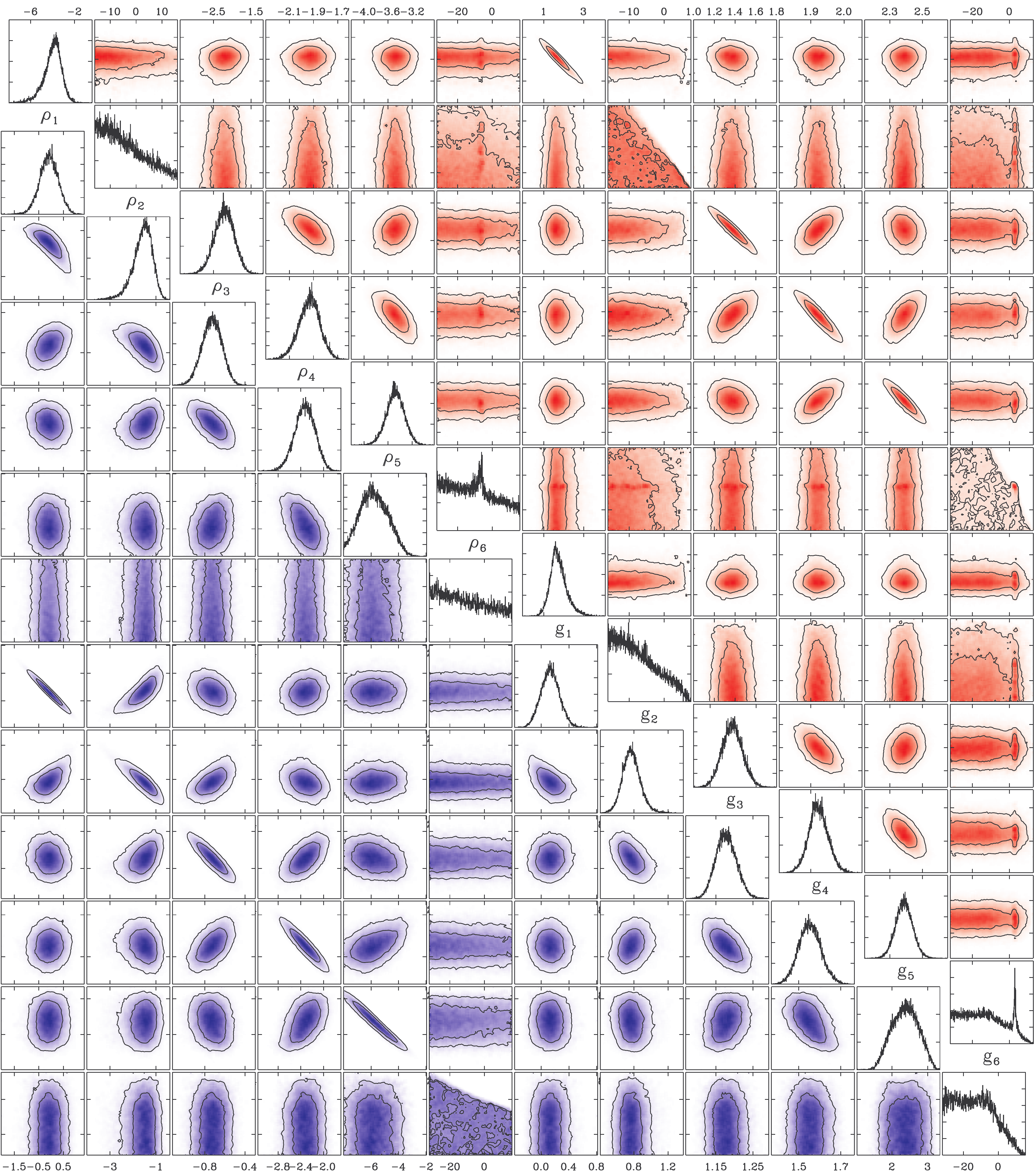}
\caption{Likelihood space of $\log(\rho_i)$ and $\log(g_i)$ for the
  two models, as sampled using MCMC. The $C$--$L$ evolution ($\alpha =
  1$) and no $C$--$L$ evolution ($\alpha = 0$) models are shown, in
  blue (lower-left) and red (upper-right), respectively. For each set,
  the diagonal shows the likelihood of each of the free parameters
  ($\rho_i$ followed by $g_i$, from top-left to bottom-right, for $z_i
  = 0.1$, 0.5, 1.0, 2.0, 3.5 and 5.0). Below/above the diagonal are
  the correlations between all pairs of parameters. Contours are the
  68 and 95 per cent confidence regions. We note that $\rho$ and $g$
  at each $z_i$ are highly anti-correlated, and there are moderate
  anti-correlations between adjacent redshift nodes. For both data
  sets, the $z=5$ nodes are not constrained, and the $z=0.5$ nodes are
  unconstrained for the $\alpha = 0$ model, as can be seen in
  Fig.~\ref{fig:params}. A spike is apparent in the $\alpha = 0$, $z=5$
  nodes; this feature appears to be consistent across and throughout
  the chains, but as it is uncorrelated with the other more
  well-constrained, lower-redshift parameters, we believe it has no
  effect on model predictions.}
\label{fig:chains}
\end{figure*}

The parameter distributions and correlations are shown in
Fig.~\ref{fig:chains}. The distributions for the $\alpha = 1$ model
are shown in the lower-left, and for the $\alpha = 0$ model in the
upper-right. The diagonal elements show the distributions of the free
parameters, $\rho_1, \dots, \rho_6, g_1, \dots, g_6$. Above and below
the diagonals are the joint distributions of all pairs of
parameters. We note a few features: (i) the $z_1 = 0.5$ parameters for
the $\alpha = 0$ model are unconstrained, as was noted above; (ii) the
$z_6 = 5.0$ parameters are unconstrained by the data for both models,
also noted above; (ii) for both $\rho$ and $g$, neighbouring
parameters (e.g., $\rho_i$ and $\rho_{i+1}$) are anti-correlated; and
(iv) $\rho_i$ and $g_i$ (at the same $z_i$) are strongly
anti-correlated.

\subsection{Comparison to Data}

Comparisons of the model to the constraining data sets are shown in
Figs.~\ref{fig:counts1}--\ref{fig:nzdist}.

\subsubsection{Number Counts}
\label{sec:rescounts}

\begin{figure*} 
\centering
\includegraphics[width=\linewidth]{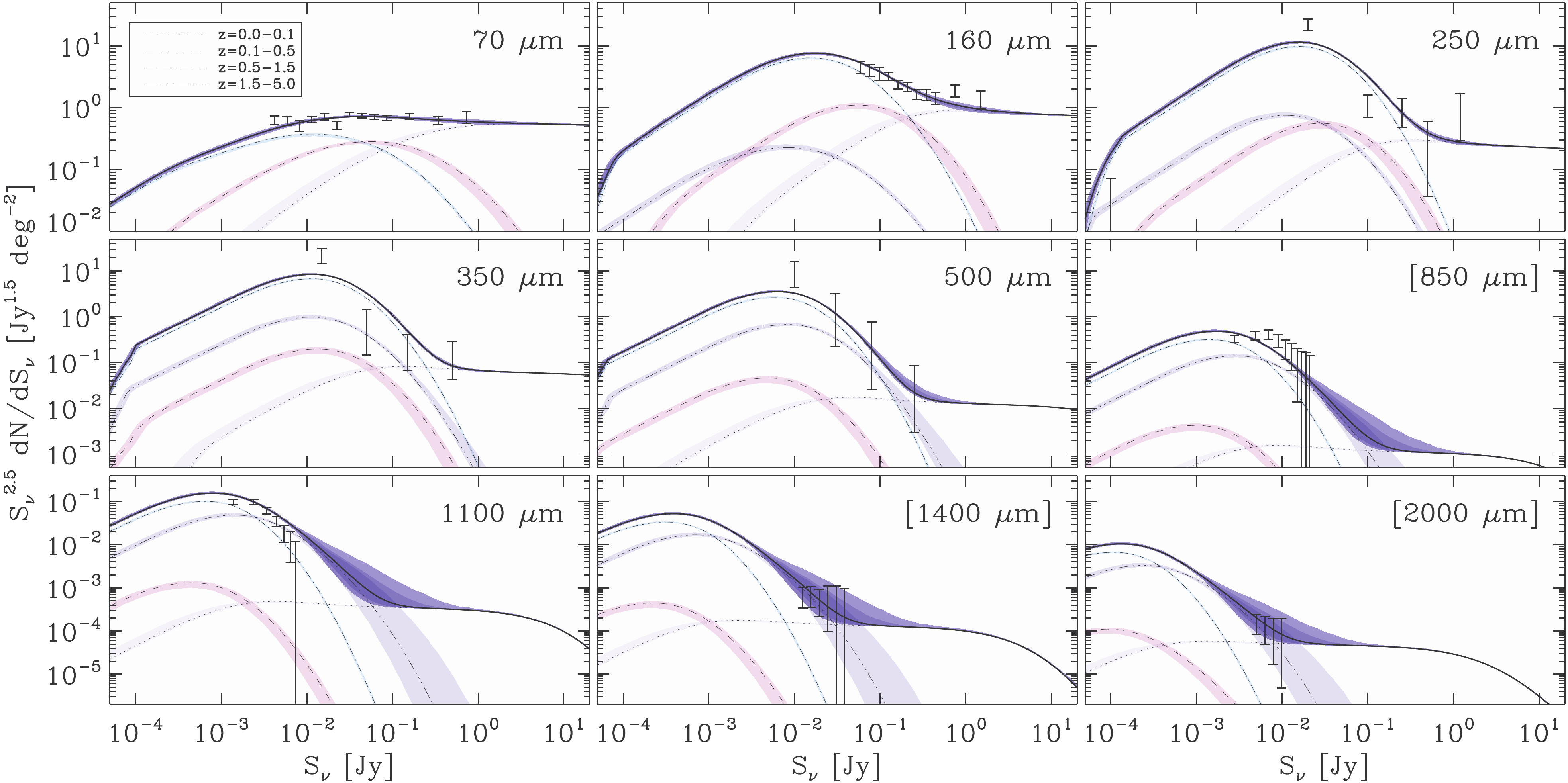}
\caption{Euclidean-normalised differential number counts. The solid
  lines are the counts derived at each wavelength from the best-fit
  $\alpha = 1$ model. The coloured bands represent the 68 and 95 per
  cent confidence regions. The contribution of sources at
  $z=0.0$--0.1, 0.1--0.5, 0.5--1.5 and 1.5--5.0 are shown as dotted,
  dashed, dot-dashed and triple-dot-dashed lines, respectively. The
  coloured bands indicate the 68 per cent confidence regions. We note
  that the $z=0.1$--0.5 component is sub-dominant at all flux
  densities and wavelengths, except for a narrow region in the
  shortest bands. The BLAST data sets (250--350\,\micron) include
  poorly-constrained points at low and high flux density that are not
  included in the range shown here; see
  Fig.~\ref{fig:intcounts}. Counts at 850\,\micron\ (SCUBA) and 1.4
  and 2.0\,mm (SPT) are shown for comparison only, and are not used in
  the fits.}
\label{fig:counts1}
\end{figure*}

\begin{figure*} 
\centering
\includegraphics[width=\linewidth]{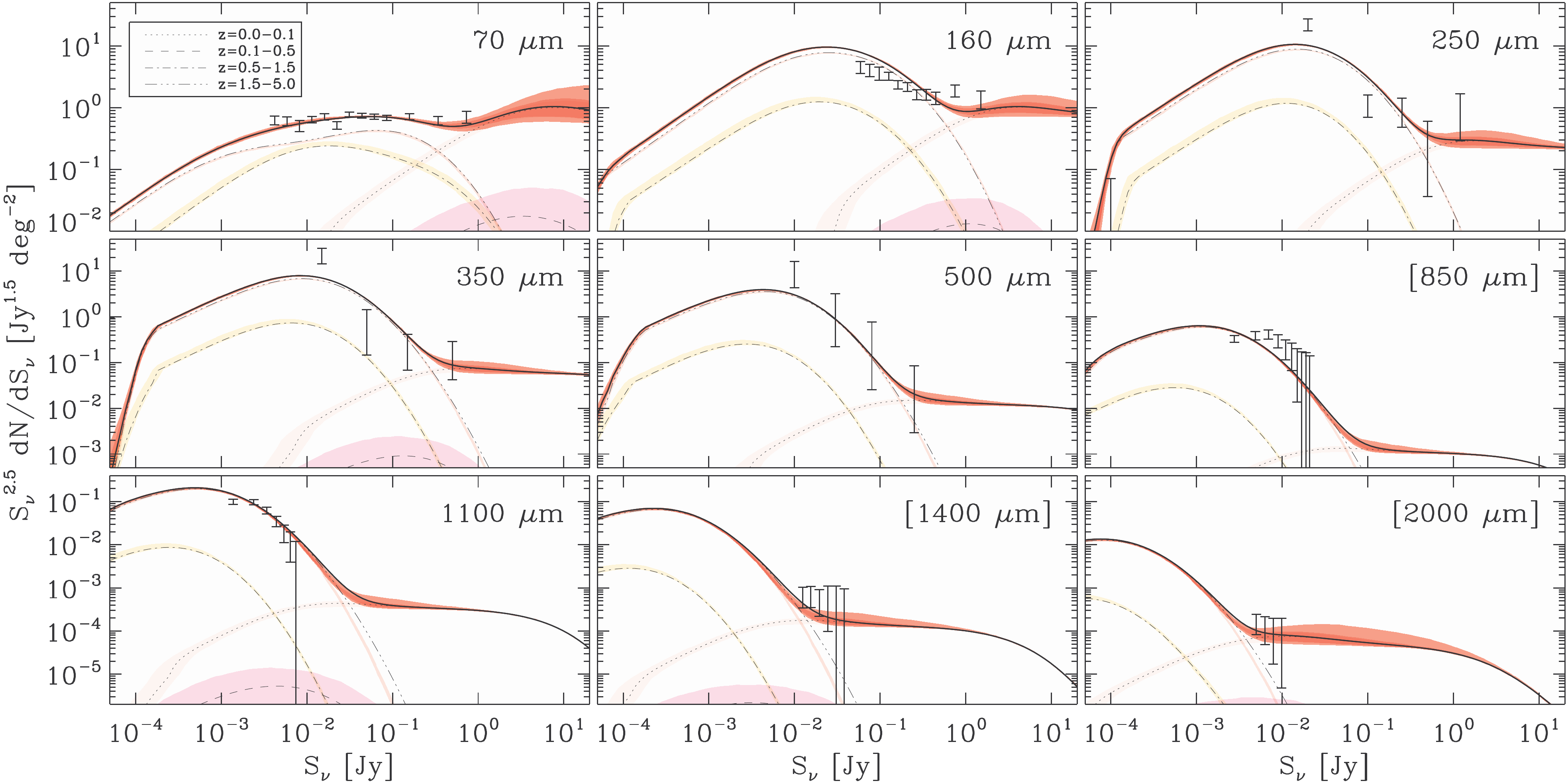}
\caption{Same as Fig.~\ref{fig:counts1}, but for $\alpha = 0$. Note
  that, with the exception of 160\,\micron, which is not well-fit, the
  shape of the $\alpha=0$ counts are very similar to the $\alpha=1$
  counts (Fig.~\ref{fig:counts1}), but the contribution from the
  various redshift ranges are completely different. In particular, the
  $z=0.1$--0.5 contribution never dominates at any band or flux
  density.}
\label{fig:counts2}
\end{figure*}

Euclidean-normalised differential number counts at a range of
wavelengths are shown in Figs.~\ref{fig:counts1} and \ref{fig:counts2}
(for the $\alpha = 1$ and $\alpha = 0$ models, respectively). The
counts derived from the best-fit models are shown as solid lines, with
the coloured regions showing the 68 and 95 per cent confidence
regions. Also shown are the contributions to the counts from sources
binned by redshift, along with the 68 per cent confidence regions. The
counts at 850, 1400 and 2000\,\micron\ are shown for illustrative
purposes only, and have not been used in the fits.

Comparing the counts derived from the $\alpha = 1$ and 0 models, we
see that, with the exception of 160\,\micron, which the $\alpha=0$
model is not able to reproduce, the two models give very similar
counts through the flux density regions covered by data. We see,
however, that the curves are formed from completely different redshift
distributions; the $\alpha=1$ model shows a prominent $z=0.1$--0.5
component at all wavelengths, while this component is completely
sub-dominant in the $\alpha=0$ model. This fact provides an
explanation for why the $z=0.5$ density and luminosity evolution
parameters in the $\alpha=0$ model are completely unconstrained -- since
sources at this redshift do not contribute appreciably to the observed
counts at any wavelength, there is only an upper limit to the strength
of sources in this redshift range.

\begin{figure} 
\centering
\includegraphics[width=\linewidth]{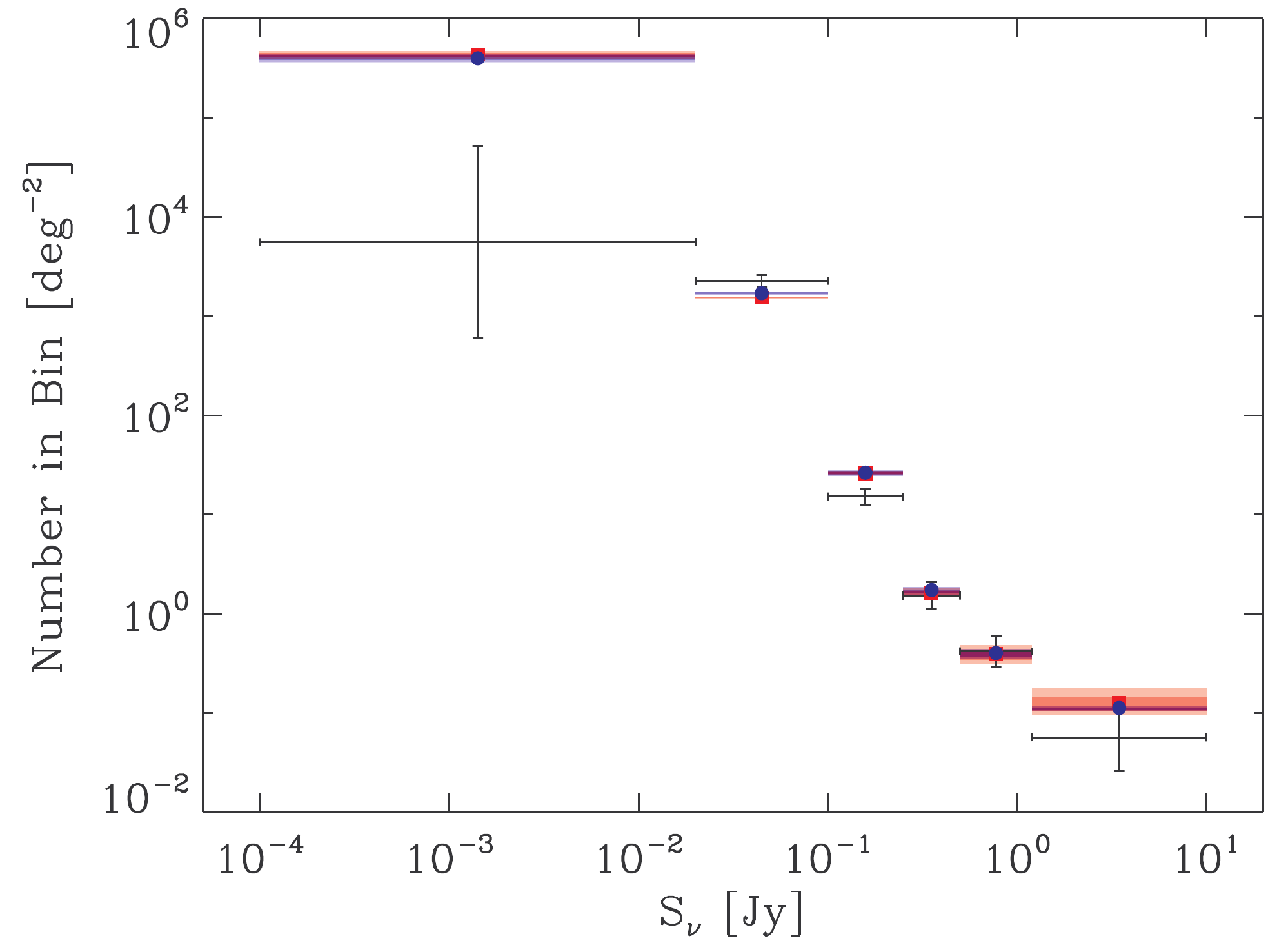}
\caption{The BLAST 250\,\micron\ $P(D)$ counts integrated between
  nodes compared to models (colour-coded as in the previous figures).
  The horizontal error bars indicate the
  size of the bins (the bin edges correspond to the flux densities of
  the nodes listed in \patanchon\ and plotted in
  Figs.~\ref{fig:counts1} and \ref{fig:counts2}), while the vertical
  bars show the errors determined from Monte Carlo sampling of the
  \patanchon\ covariance matrix.}
\label{fig:intcounts}
\end{figure}

We remind the reader that the 250, 350 and 500\,\micron\ BLAST counts
are fit using the integral of $dN/dS_\nu$ across a set of bins. An
example fit (250\,\micron\ for the $\alpha = 1$ model) is presented in
Fig.~\ref{fig:intcounts}. This shows very clearly that the two models
are able to produce essentially the same counts. We also point out that the
bin-to-bin correlations are in some cases quite strong (up to $\rho =
0.95$), and that the $\chi^2$ calculated from the full covariance
matrix can be quite different from what would be inferred using the
diagonal elements alone. This is particularly true at 500\,\micron,
where $\chi^2 = 45$; if we ignore the correlations, we instead find
$\chi^2 = 7$.

Although best-fits have been found, it is worth pointing out that none of
the overall fits are formally `good'; in particular, the model is quite low
at the $S_\nu=10-20$\,mJy BLAST counts and is high compared to
1100\,\micron\ counts at the bright end. We believe that, due to the choice
of parameterisation in the \patanchon\ $P(D)$ counts, these points are
biased high. This comes about since a connected power law is not able
to reproduce the curvature displayed by the model. Recent results from
\textsl{Herschel\/}/SPIRE using $P(D)$ to measure counts
\citep{glenn2010}, which go to fainter flux densities than the BLAST
counts, are low compared to BLAST, supporting this hypothesis (which
we discuss further at the end of Section~\ref{sec:goodnessoffit}).

Another assumption that may be biasing the fits is that the 3 BLAST
measurements are correlated with one-another, both by instrumental
noise and by the fact the same part of the sky has been
observed. For both of these reasons, it would be desirable to fit the
model directly to the maps via multi-band $P(D)$. This will be the
focus of a later paper.

\subsubsection{Integrated Brightness (CIB)}

\begin{figure} 
\centering
\includegraphics[width=\linewidth]{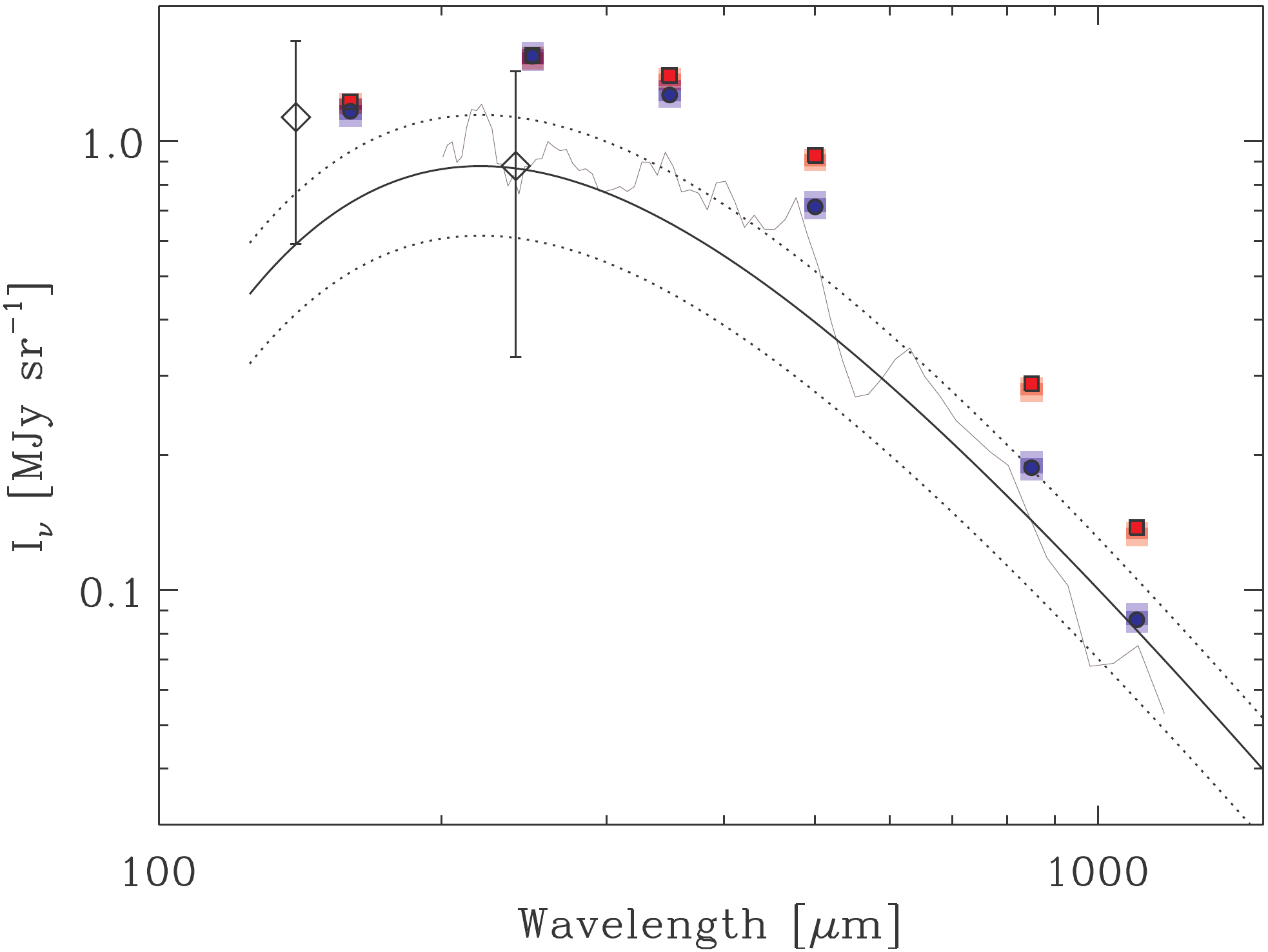}
\caption{The cosmic infrared background as measured by FIRAS
  \citep[][solid line]{fixsen1998} with representative 30 per cent
  errors (dotted lines). The circular and square points represent the
  CIB derived from the best-fit models, $\alpha=1$ and $\alpha=0$,
  respectively . The coloured rectangles represent the 68 and 95 per
  cent confidence regions. Also shown, but not used in the fits, are
  the FIRAS (solid grey line) and DIRBE/WHAM measurements (diamonds)
  by \citet{lagache2000}.  Our models tend to over-predict the background
  values; we discuss the reason for this in Section~\ref{sec:discussion}.}
\label{fig:cib}
\end{figure}

The CIB as measured by FIRAS is used as a constraint on the integral
of intensity over redshift at a range of wavelengths. This is shown in
Fig.~\ref{fig:cib} for both models. The best-fit models are shown as
filled symbols, with the 68 and 95 per cent confidence regions shown
as coloured rectangles. We see that both models are high compared to
FIRAS at nearly all wavelengths. It may be interesting to note that
the models appear to agree with the \citet{lagache2000} curve slightly
better than the \citet{fixsen1998} curve.

As with the correlations between bands in the BLAST counts, we have
also ignored correlations between bands in the FIRAS
measurement. Proper treatment of these correlations would likely
reduce the overall $\chi^2$ of the models.

\subsubsection{Redshift Distribution}

\begin{figure} 
\centering
\includegraphics[width=\linewidth]{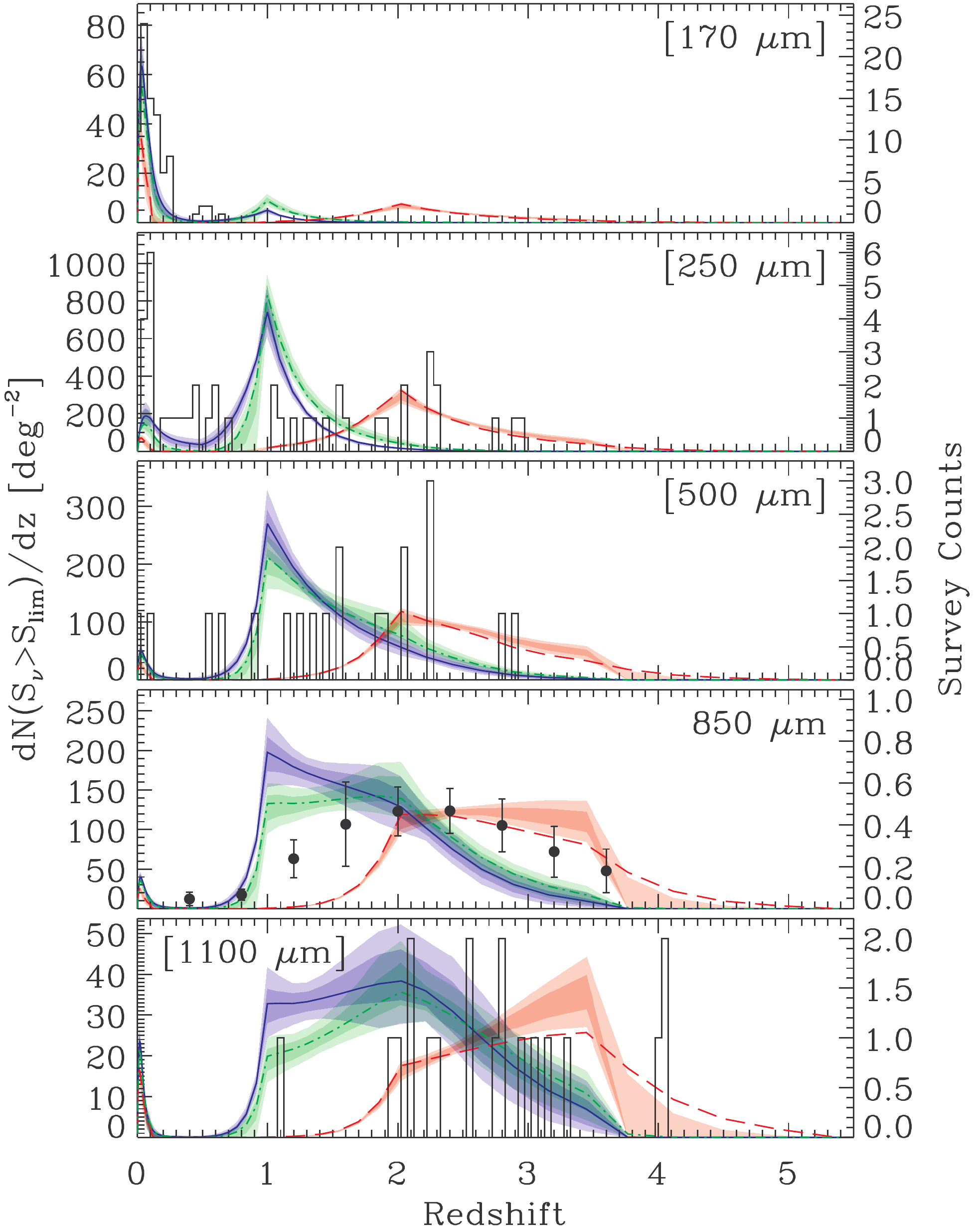}
\caption{The redshift distribution of number counts above a flux
  density limit. The thick solid/long-dashed/dot-dashed curves show
  the predictions of the best-fit $\alpha=1$/$\alpha=0$/free $\alpha$
  models. The coloured bands represent the 68 and 95 per cent
  confidence regions. At 170\,\micron\ (\textsl{ISO\/}), 250 and
  500\,\micron\ (BLAST) and 1.1\,mm (AzTEC), redshift histograms are
  shown. At 850\,\micron\ (SCUBA), we show the scaled histogram of
  \citet{chapman2005} with Poisson error bars, as described in
  Section~\ref{sec:reddist}. Redshift distributions at 170
  \citep{patris2003,dennefeld2005,taylor2005}, 250,
  500\,\micron\ \citep{chapin2011} and 1.1\,mm \citep{chapin2009b} are
  shown for comparison only, and are not used in the fits.}
\label{fig:nzdist}
\end{figure}

In Fig.~\ref{fig:nzdist}, we show a variety of $dN/dz$ measurements:
170\,\micron\ from \textsl{ISO\/}; 250 and 500\,\micron\ from BLAST;
850\,\micron\ from SCUBA; and 1.1\,\micron\ from AzTEC. As discussed
in Section~\ref{sec:datanotused}, the selection function for these
redshift distributions are very poorly quantified, so we do not make
full use of these counts as constraints on the model. However, to
provide at least some direct redshift constraint, we use the approximated
SCUBA distribution, and at the other wavelengths simply show the
counts predicted by the model compared to the data.

\subsection{Goodness-of-fit}
\label{sec:goodnessoffit}

Table~\ref{tab:results} lists the best-fit $\chi^2$, along with the
contribution from each data set, for the $\alpha = 1$, $\alpha = 0$
and the free $\alpha$ models (columns 3, 9 and 10, respectively). The
number of data points in each data set is listed in column 2.  The
total $\chi^2$ and number of degrees of freedom (DOF) for each fit are
listed along the bottom two rows. Since there are 62 data points and
12 (13) free parameters, there are 50 (49) DOF for the fixed (free)
$\alpha$ models. The reduced $\chi^2$, $\chi^2_\mathrm{r} = \chi^2 /
N_\mathrm{dof}$, is an unfortunately high 3.8, 6.8 and 3.5 for the
$\alpha=1$, $\alpha=0$ and free $\alpha$ models,
respectively. Clearly, the $\alpha=1$ model is a much better fit to
the data than the $\alpha=0$ model (but see
Section~\ref{sec:reconciling}). Adding an extra parameter for the free
$\alpha$ model also appears to be justified; even so, we focus on the
$\alpha=1$ model, to keep things simple.

\begin{table*}
\caption{Breakdown of $\chi^2$ by data set for different models. The
  total $\chi^2$ and number of degrees of freedom, $N_{\rm dof}$, are listed in
  the bottom two rows. In columns 4--8, the $\alpha=1$ model is re-fit
  excluding (or replacing) individual data sets. The $\chi^2$ for
  excluded data sets are listed in parentheses.}
\label{tab:results}
\begin{tabular}{cccccccccc}
\hline
\textbf{BAND} & 
\textbf{ No.~Pts} &
\multicolumn{6}{c}{\myhrulefill\ \textbf{{\boldmath $\alpha = 1$}} \myhrulefill} &
\textbf{\boldmath $\alpha = 0$} &
\textbf{{\boldmath $\alpha$} free}
\\
\textbf{({\boldmath $\mu$}m)} & 
& 
\textbf{All} & 
\textbf{No 250 and} & 
\textbf{No 1100} &
\textbf{850} &
\textbf{No {\boldmath $dN/dz$}} & 
\textbf{No CIB} &
\textbf{All} & 
\textbf{All} \\
\textbf{data} &
&
&
\textbf{500 counts} &
\textbf{counts} &
\textbf{counts} &
& & 
\textbf{data} & 
\textbf{data} \\
\hline \hline
\multicolumn{10}{c}{\textit{Counts}} \\
70   & 
13 & 
\phantom{0(}12.2\phantom{)}  &  
\phantom{0(}14.4\phantom{)}  &   
\phantom{0(}13.6\phantom{)}  &
\phantom{0(}13.5\phantom{)}  &
\phantom{0(}14.3\phantom{)}  &  
\phantom{0}10.1 & 
\phantom{0}15.6 &  
\phantom{0}14.6  \\
160  & 
11 & 
\phantom{00(}6.5\phantom{)}  &  
\phantom{0(}13.9\phantom{)}  &    
\phantom{00(}7.7\phantom{)}  &    
\phantom{00(}8.4\phantom{)}  &    
\phantom{00(}8.9\phantom{)}  &   
\phantom{00}5.0 & 
\phantom{0}71.9 &   
\phantom{00}8.3  \\
250  & 
\phantom{0}6 & 
\phantom{0(}10.0\phantom{)}  & 
\phantom{0}(28.0) &   
\phantom{0(}13.8\phantom{)}  &   
\phantom{0(}10.7\phantom{)}  &   
\phantom{0(}13.1\phantom{)}  &  
\phantom{0}11.2 & 
\phantom{0}19.5 &  
\phantom{0}14.0  \\
350  & 
\phantom{0}5 & 
\phantom{0(}16.7\phantom{)}  &  
\phantom{0(}20.5\phantom{)}  &   
\phantom{0(}17.3\phantom{)}  &   
\phantom{0(}17.8\phantom{)}  &   
\phantom{0(}15.9\phantom{)}  &   
\phantom{00}7.7 & 
\phantom{0}47.5 &  
\phantom{0}18.5  \\
500  & 
\phantom{0}5 & 
\phantom{0(}45.3\phantom{)}  & 
\phantom{0}(72.7) &   
\phantom{0(}13.8\phantom{)}  &   
\phantom{0(}25.6\phantom{)}  &   
\phantom{0(}31.5\phantom{)}  &  
\phantom{0}28.5 & 
\phantom{0}81.8 &  
\phantom{0}48.1  \\
850 &
10 &
\phantom{0}(13.1) &
\phantom{0}(22.1) &
\phantom{0}(49.1) &   
\phantom{0(}13.7\phantom{)} & 
\phantom{0}(14.6)  &
\phantom{0}(18.8)  &   
\phantom{0}(20.0)  &
\phantom{0}(14.8)  \\
1100 & 
\phantom{0}7 & 
\phantom{0(}31.6\phantom{)} &   
\phantom{00(}7.8\phantom{)} & 
(916.9)  &
(359.0)  &   
\phantom{0(}12.0\phantom{)}  &  
\phantom{0}26.1 & 
\phantom{0}19.8 &  
\phantom{0}24.7  \\
\hline
\multicolumn{10}{c}{\textit{Background}} \\
 160 &  
\phantom{0}1 &  
\phantom{00(}4.0\phantom{)}  &   
\phantom{00(}2.5\phantom{)}  &    
\phantom{00(}1.7\phantom{)}  &    
\phantom{00(}2.5\phantom{)}  &    
\phantom{00(}1.8\phantom{)}  & 
(3.2$\times 10^3$)   &  
\phantom{00}5.1 &   
\phantom{00}1.9  \\
 250 &  
\phantom{0}1 &  
\phantom{00(}7.2\phantom{)}  &   
\phantom{00(}4.0\phantom{)}  &    
\phantom{00(}4.0\phantom{)}  &    
\phantom{00(}4.7\phantom{)}  &    
\phantom{00(}4.3\phantom{)}  & 
(3.9$\times 10^5$)       &  
\phantom{00}7.3 &   
\phantom{00}4.6  \\
 350 &  
\phantom{0}1 &  
\phantom{00(}9.6\phantom{)}  &   
\phantom{00(}3.9\phantom{)}  &    
\phantom{00(}7.2\phantom{)}  &    
\phantom{00(}6.9\phantom{)}  &    
\phantom{00(}6.7\phantom{)}  & 
(2.7$\times 10^6$)       & 
\phantom{0}14.2 &   
\phantom{00}7.7  \\
 500 &  
\phantom{0}1 &  
\phantom{00(}7.2\phantom{)}  &   
\phantom{00(}2.0\phantom{)}  &    
\phantom{00(}7.8\phantom{)}  &    
\phantom{00(}5.8\phantom{)}  &    
\phantom{00(}4.6\phantom{)}  & 
(1.7$\times 10^7$)       & 
\phantom{0}20.4 &   
\phantom{00}6.4  \\
 850 &  
\phantom{0}1 &  
\phantom{00(}1.1\phantom{)}  &   
\phantom{00(}0.0\phantom{)}  &    
\phantom{00(}2.9\phantom{)}  &    
\phantom{00(}1.1\phantom{)}  &    
\phantom{00(}0.2\phantom{)}  & 
(2.4$\times 10^7$)       & 
\phantom{0}11.4 &   
\phantom{00}1.1  \\
1100 &  
\phantom{0}1 &  
\phantom{00(}0.0\phantom{)}  &  
\phantom{00(}0.4\phantom{)}  &    
\phantom{00(}0.8\phantom{)}  &    
\phantom{00(}0.1\phantom{)}  &    
\phantom{00(}0.1\phantom{)}  & 
(3.6$\times 10^6$)       &  
\phantom{00}5.3 &   
\phantom{00}0.1  \\
\hline
\multicolumn{10}{c}{$dN/dz$} \\
 850 &  
\phantom{0}9 & 
\phantom{0(}40.8\phantom{)}  &  
\phantom{0(}52.8\phantom{)}  & 
\phantom{00(}3.3\phantom{)}  & 
\phantom{00(}7.2\phantom{)}  & 
(238.5) &  
\phantom{0}24.1 & 
\phantom{0}20.1 &  
\phantom{0}20.8  \\
\hline \hline
\multicolumn{2}{c}{\textbf{ Total:}} & 
\phantom{(}192.3\phantom{)} & 
\phantom{(}122.2\phantom{)} & 
\phantom{0(}93.9\phantom{)} &
\phantom{(}117.9\phantom{)} &
\phantom{(}113.4\phantom{)} & 
112.7 & 
339.8 & 
170.6 \\
\multicolumn{2}{c}{\textbf{ $N_{\rm dof}$:}} & 
\phantom{0(}50\phantom{.0)} & 
\phantom{0(}39\phantom{.0)} & 
\phantom{0(}43\phantom{.0)} &
\phantom{0(}53\phantom{.0)} &
\phantom{0(}41\phantom{.0)} & 
\phantom{0}44\phantom{.0} & 
\phantom{0}50\phantom{.0} & 
\phantom{0}49\phantom{.0} \\
\multicolumn{2}{c}{\textbf{\boldmath $\chi^2_\mathrm{r}$:}} & 
\phantom{00(}3.8\phantom{)} & 
\phantom{00(}3.1\phantom{)} & 
\phantom{00(}2.2\phantom{)} &
\phantom{00(}2.2\phantom{)} &
\phantom{00(}2.8\phantom{)} & 
\phantom{00}2.6 & 
\phantom{00}6.8 & 
\phantom{00}3.5 \\
\hline
\end{tabular}
\end{table*}

We explore the effects of re-fitting the model omitting various data
sets in order to probe the `strain' on the model due to any particular
data set.  We have run the $\alpha=1$ model on all data sets: (i)
excluding 250 and 500\,\micron\ counts; (ii) excluding
1100\,\micron\ counts; (iii) using 850\,\micron\ counts \emph{instead
  of\/} 1100\,\micron\ counts; (iv) excluding the 850\,\micron\ $dN/dz$
distribution; and (v) excluding the CIB. The results of these tests are
discussed here:
\begin{enumerate}
\item The first test (column 4) was meant to explore the effects of
  correlations between the BLAST data sets; however, it is not clear
  how much of the improvement in $\chi^2_\mathrm{r}$ is due to
  correlations and how much is due to the lessening of tension between
  BLAST and 1100\,\micron. The value of $\chi^2_\mathrm{r}$ is slightly lower,
  at 3.1, and we see that the 1100\,\micron\ counts and CIB are fit much
  better, although the $dN/dz$ agreement is much worse.

\item Removing the 1100\,\micron\ counts (column 5) greatly increases
  the goodness-of-fit, reducing $\chi^2_\mathrm{r}$ to 2.2. We see
  that the tension between the BLAST and 1100\,\micron\ data sets is
  greatly relieved, that the background is reduced at the shorter
  wavelengths, and that $dN/dz$ is allowed to fit nearly perfectly.
  We believe this is a strong clue for developing improved models, as we
  discuss in the next section.

\item Fitting the 850\,\micron\ counts instead of the
  1100\,\micron\ counts (column 6) greatly improves the fit compared to the
  full data set, to a reduced $\chi^2_\mathrm{r}$ to 2.2. This is because,
  compared to 1100\,\micron\ counts, the 850\,\micron\ counts are
  higher at the bright end and lower at the faint end, which allows
  better fits to the 500 and 850\,\micron\ counts. The
  850\,\micron\ $dN/dz$ distribution is allowed to fit well. We note,
  however, that the model significantly over-predicts the 1.4 and
  2.0\,mm counts (not shown here).

\item Removing the 850\,\micron\ $dN/dz$ constraint (column 7) also
  removes tension, in this case between BLAST and
  1100\,\micron\ counts, although not to such a high degree as for
  (ii); here, $\chi^2_\mathrm{r} = 2.8$. 

\item Without the CIB constraints (column 8), we see that the
  integrated background is entirely unbounded. This is because the CIB
  is the only available constraint on the faint end of the counts,
  which dominates the CIB if the faint-end counts are sufficiently
  steep. The quality of fit to all other data set is improved, with
  $\chi^2_\mathrm{r} = 2.6$. However, with no overall constraint at
  faint flux densities, the number of high-redshift sources is greatly
  increased. We note that this is not reflected in the
  850\,\micron\ $dN/dz$ constraint, since that data set includes only
  galaxies brighter than $S_{850} > 5$\,mJy.
\end{enumerate}

We have also run a fit of the $\alpha = 1$ model to test the
hypothesis that the low-flux density BLAST nodes are biased high by
the $P(D)$ parameterisation, as discussed in
Section~\ref{sec:rescounts}. We have adjusted the 20/15/10\,mJy nodes
of the 250/350/500\,\micron\ counts to match the \citet{glenn2010}
\textsl{Herschel\/}/SPIRE $P(D)$ counts, and have also adjusted the
values of the other nodes based on the BLAST counts covariance
matrices. Additionally, we have doubled the errors on the lowest and
highest flux density nodes, to compensate for the non-Gaussian error
distributions (faint nodes) and cosmic variance (bright nodes). We
find that, with this adjusted data set, $\chi^2$ for the model is
80.3, with $\chi^2_\mathrm{r} = 1.6$. This model is a much better fit
to the 350, 500 and 1100\,\micron\ counts, the 850\,\micron\ redshift
distribution, and CIB through the BLAST bands (although the model is
still $\sim 1.5\,\sigma$ high). This test indicates that much of the
tension in the model is due to the $P(D)$ parameterisation, although
this alone is not the sole cause of the over-predicted CIB.

 To further test how much of this improvement is due to the
 `adjustment' and how much is due to the doubled error bars, we also
 ran a test with the original BLAST counts, but with the errors on the
 lowest and highest flux density nodes doubled. For this test, we find
 $\chi^2=118$. This shows that a large fraction of the tension between
 data sets could be due to the apparently high data points at low flux
 density in the BLAST $P(D)$ counts, although underestimates of the
 errors at low and high flux density (possibly due to non-Gaussian
 error distributions and cosmic variance, respectively) may also be a
 significant factor.

\section{Discussion}
\label{sec:discussion}

We now look at some inferences that can be drawn from the model,
discuss the implications of the colour-luminosity evolution degeneracy
and data sets needed to resolve it, and consider other possible
improvements, including the use of new data sets and modifications to the
techniques.

\subsection{Colour-colour distributions}
\label{sec:colcol}

In Section~\ref{sec:seds} it was shown that our SED library is
consistent with the real spread in galaxy SEDs in the local Universe
by using them to map our \textsl{IRAS\/}-based $\Phi(L,C)$ to the
independent monochromatic luminosity functions at 12, 15, 25 and
850\,\micron. However, it is possible that our chosen SED shapes could
conspire to produce these consistent \emph{integral\/} quantities, while
failing for individual galaxies. Now that we have an evolving model of
the luminosity function in hand, we can go back to our best-fit
distribution, apply observational selection functions, and compare the
distribution of colours for \emph{individual\/} galaxies from our model
to those detected in real surveys (i.e., to verify the correlations
between bands). For this test, we have relied on the two best examples
that we could find in the literature of surveys with colour
information spanning 24--850\,\micron: an $S_{70} > 5$\,mJy \textsl{Spitzer\/}
survey with cross-matched 24 and 160\,\micron\ observations taken as
part of the Cosmological Evolution Survey
\citep[COSMOS,][]{kartaltepe2010}; and the Submillimetre Local
Universe Galaxy Survey (SLUGS), in which SCUBA was used to follow up
$S_{60} > 5.24$\,Jy \textsl{IRAS\/} galaxies at 450 and 850\,\micron\
\citep{dunne2000,dunne2001}.  Almost all of the SLUGS galaxies
have luminosities $L > 10^{10}$\,L$_\odot$, with about half above $L >
10^{11}$\,L$_\odot$, and it is a truly local sample, with all of the
objects lying at $z<0.1$.  The COSMOS sample is deeper and
higher-redshift, though with a similar range of luminosities; about
50\% of the sources lie at $z>0.5$, and virtually all of the objects
have luminosities $L > 10^{10}$\,L$_\odot$, while $\sim70\%$ have
luminosities $L > 10^{11}$\,L$_\odot$.

\begin{figure}
\centering
\includegraphics[width=\linewidth]{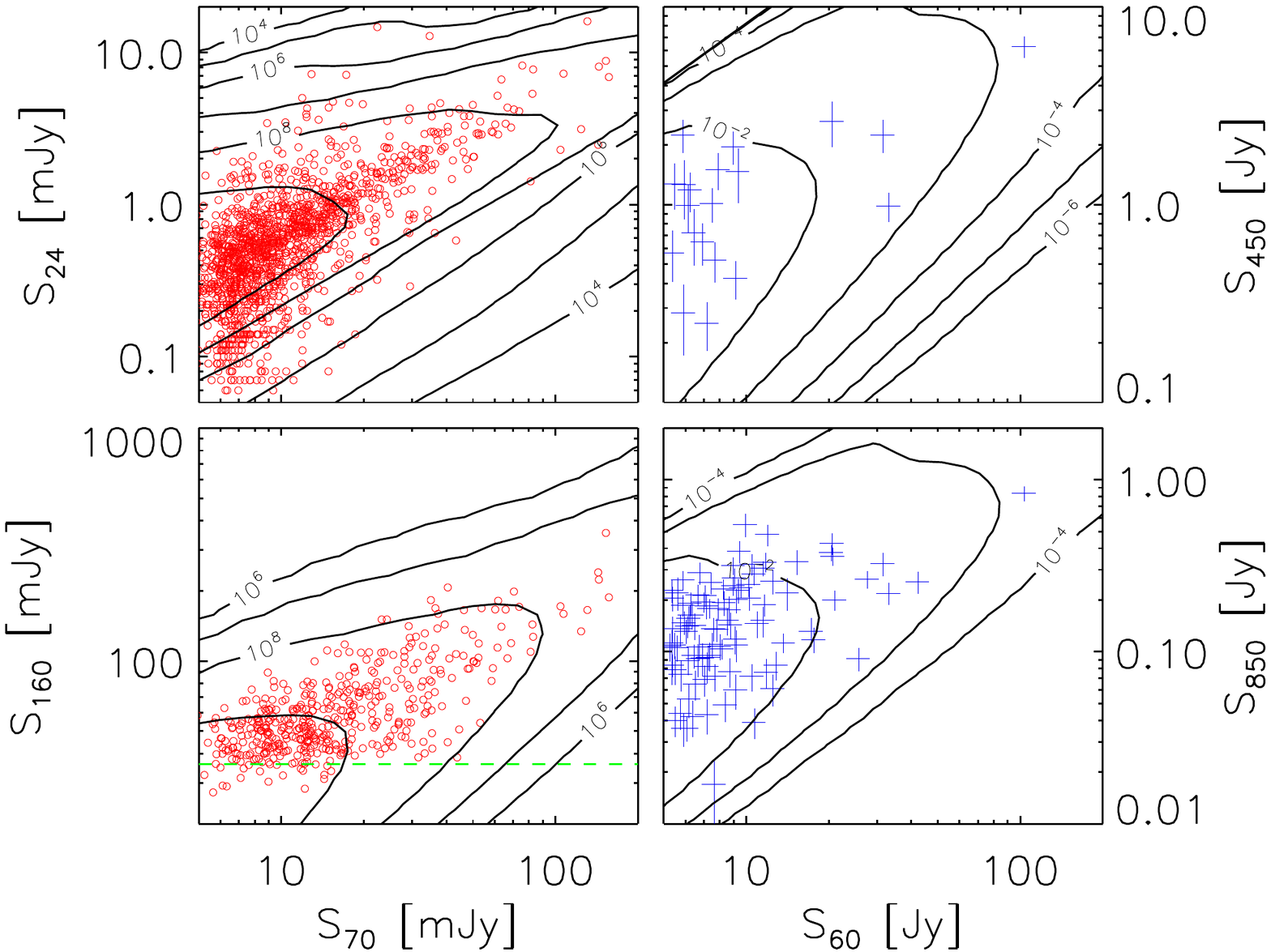}
\caption{Comparison between the colours of galaxies drawn from our
  best-fit evolving luminosity function (contours) with data from real
  surveys (symbols). The left panels are galaxies selected at
  70\,micron\ above 5\,mJy \citep[red symbols,][]{kartaltepe2010}. The
  top-left panel shows 24\,\micron\ vs.\ 70\,\micron\ and the
  bottom-left panel shows 160\,\micron\ vs.\ 70\,\micron. There is an
  approximate flux-limit of 36\,mJy in the 160\,\micron\ data, which is
  indicated by a green dashed line. The right panels compare our model
  with SCUBA 450 and 850\,\micron\ follow-up of 60\,\micron\ sources
  brighter than 5.24\,Jy \citep[blue symbols whose width and height
    indicate measurement errors,][]{dunne2000,dunne2001}. The top
  right panel shows 450\,\micron\ vs.\ 60\,\micron, and the bottom
  right panel shows 850\,\micron\ vs.\ 60\,\micron. In all panels, the
  contours indicate the density of sources predicted from the model,
  with units of $(\log \mathrm{Jy})^{-2} \, \mathrm{deg}^{-2}$. Unlike
  Fig.~\ref{fig:locallumfunc}, which shows that our SED library and
  local colour-luminosity distribution can produce the correct
  \emph{total numbers\/} of galaxies in several bands, this comparison
  validates our model for \emph{individual objects\/}.}
\label{fig:colcol}
\end{figure}

The comparison of our model to these data sets
is shown in Fig.~\ref{fig:colcol}. The \textsl{Spitzer\/}
COSMOS data shows excellent correspondence with our model in terms of
the 160-to-70\,\micron\ colour.  Our theoretical distribution also
broadly reproduces the main trend in the observed 24-to-70\,\micron\
correlation, although the scatter is slightly less than that observed
in real galaxies. As stated in the introduction, we have \emph{not\/}
attempted to fully reproduce the observed properties of 24\,\micron\
sources, since the SEDs of galaxies at those shorter wavelengths depend
on other factors (which would then require more variables). We also
note that since this is a higher-redshift sample, the 24\,\micron\
band will sample the PAH features for a small subset. The
correspondence with the SLUGS galaxies is excellent, showing that the
spread and correlation between the Rayleigh-Jeans side of our SED
templates and the FIR peak is accurate.

\subsection{Star Formation Rate History}

\begin{figure}
\centering
\includegraphics[width=\linewidth]{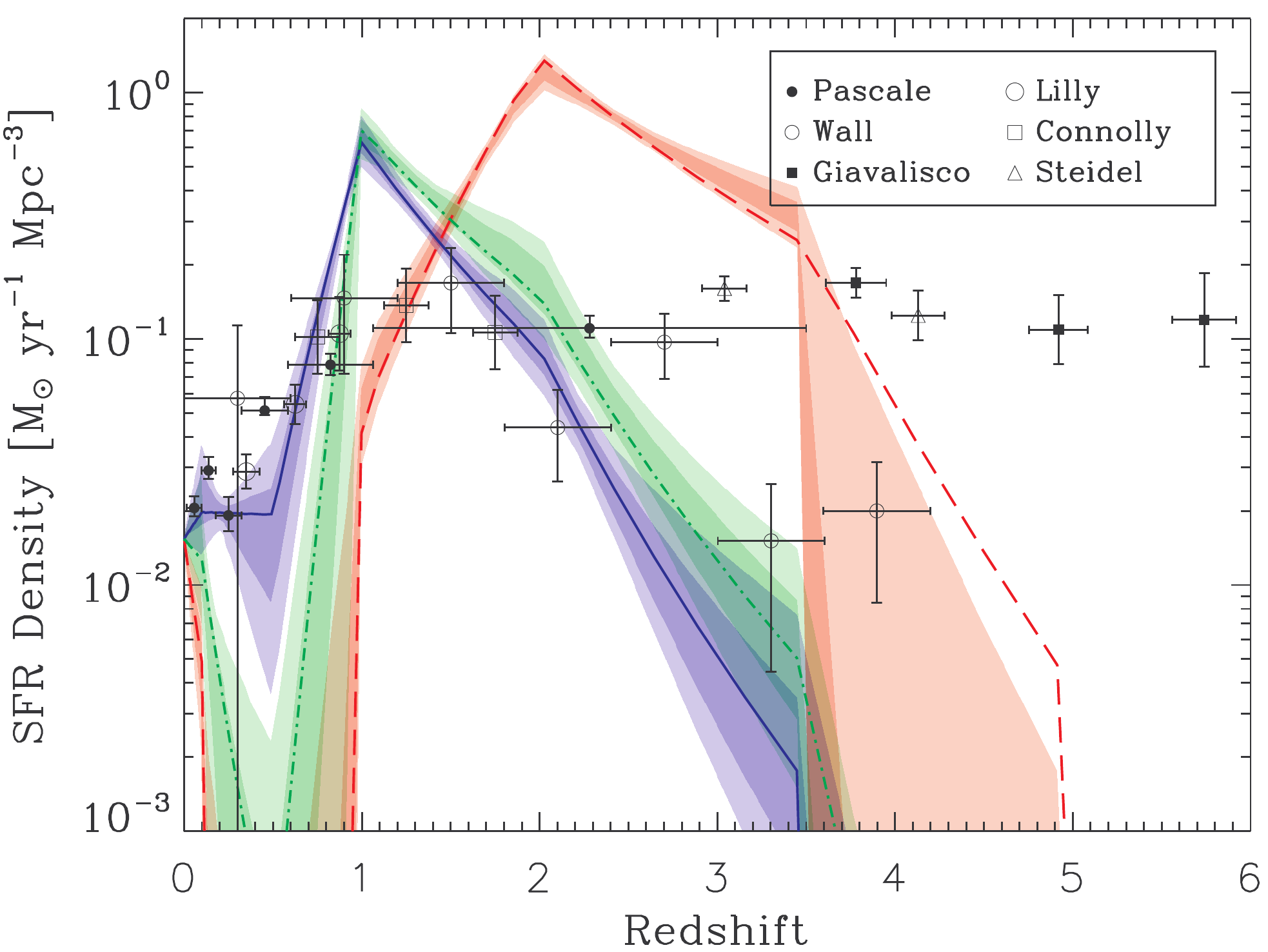}
\caption{Star formation rate density as a function of redshift. The
  $\alpha = 1$ (blue), $\alpha = 0$ (red) and free$-\alpha$ (green)
  models are shown. A sample of other estimated data points are shown;
  data are from \citet{pascale2009}, \citet{wall2008},
  \citet{giavalisco2004}, \citet{lilly1996}, \citet{connolly1997} and
  \citet{steidel1999} \citep[the latter four taken from][which has an
    extensive list of values in its appendix]{michalowski2010}.  Note
  that the jagged appearance in the model predictions is partly a
  result of the anti-correlations between adjacent redshift nodes in
  the fits (as seen in Fig.~\ref{fig:chains}).}
\label{fig:madau}
\end{figure}

Star formation rate density (SFRD) can be derived from the model by
calculating the luminosity density $L_\mathrm{TIR}(z)$:
\begin{equation}
L_\mathrm{TIR}(z) = \int_{-\infty}^{\infty} dL \int_{-\infty}^{\infty} dC\, \left[ L \,\frac{dL_\mathrm{TIR}}{dL}(C) \right] \, \Phi(L,C,z),
\end{equation}
where $L_\mathrm{TIR}$ is the luminosity integrated over the total IR range,
8--1000\,\micron, and $dL_\mathrm{TIR}/dL$, which depends only on $C$,
converts $L$ to $L_\mathrm{TIR}$. This can then be converted to SFRD
using the simple relation given by \citet{kennicutt1998},
\begin{equation}
\mathrm{SFR}\,[\mathrm{M}_\odot \,\mathrm{yr}^{-1}] = 1.728 \times 10^{-10} L_\mathrm{TIR}\,[\mathrm{L}_{\odot}].
\end{equation}
The resulting SFRD for the two models are shown in
Fig.~\ref{fig:madau}. The shapes are very different, with the $\alpha
= 1$ model peaking at $z=1$ and the $\alpha = 0$ peaking at $z=2$, but
with no SFRD whatsoever at $z<1$. We believe that this is due to the
data requiring a certain amount of cool-type galaxies; this can either
come from intrinsically cool SEDs or from redshifting
moderate-temperature SEDs so that they appear cooler. The $\alpha = 1$
model provides a higher fraction of cool galaxies with increasing
redshift, while the $\alpha = 0$ model is required to compensate by
strongly increasing the number of galaxies at high redshift; it is
then necessary to \emph{decrease\/} the number of galaxies at
low-to-moderate redshift. The free $\alpha$ model is also
shown (in green, dash-dotted line). Since the best-fit $\alpha$ value
is ${\sim}\,0.6$, it is not surprising that the SFRD curve falls
between the $\alpha=1$ and 0 models.

We show a small sampling of data along with the models. Below
$z\,{\sim}\,2$, the data are generally consistent with each other
and are broadly in line with
the $\alpha = 1$ model; this model in particular agrees well with the
\cite{wall2008} compilation.  Above $z\,{\sim}\,2$, the data are
inconsistent and it is hard to draw any firm conclusions, although we
note that the \citet{steidel1999} and \citet{giavalisco2004} points
are based on extinction-corrected UV measurements and do not
necessarily bear any relation to SFRD estimates in the FIR. The
$\alpha=0$ model appears to be inconsistent with the data. We also
note that the model with the `adjusted' BLAST counts, as described at
the end of Section~\ref{sec:goodnessoffit}, shows an SFRD curve that
is slightly higher at $z=0.5$, slightly lower at $z=1$, and runs a bit
flatter between $z=2$ and 3.5 compared to the displayed $\alpha=1$
model, fitting the \citet{wall2008} compilation slightly better.

\subsection{Fraction of CIB Resolved By BLAST}

If the CIB is in fact as high as indicated by the models
(${\sim}\,2\,\sigma$ high compared to FIRAS at the BLAST wavelengths
for the $\alpha=1$ model, Fig.~\ref{fig:cib}), the fraction of the CIB
resolved by stacking the BLAST maps on 24\,\micron\ \textsl{Spitzer\/}
sources would in fact be lower than is quoted by
\citet{marsden2009}. Assuming the CIB values predicted by the $\alpha
= 1$ model, $I_\nu=1.5$, 1.3 and 0.7\,MJy\,sr$^{-1}$, the stacked
intensities measured by \citet{marsden2009} correspond to 45, 45 and
50 per cent of the CIB, respectively. We note, however, that the CIB
predicted by the model is likely driven high by the apparently high
BLAST nodes, discussed in Section~\ref{sec:rescounts}. Comparing to
the model with the modified BLAST points, we find 55, 55 and 65 per
cent, still significantly lower than the percentages quoted in
\citet{marsden2009}, due to the fact that the model CIB is still $\sim
1.5\,\sigma$ high through the BLAST bands.

\subsection{Colour-Luminosity Evolution}
\label{sec:clevol}

We have shown that, based on counts alone, the $\alpha = 1$ and
$\alpha = 0$ models are nearly indistinguishable. The redshift
distributions, however, are very different, with the $\alpha=0$ model
a better fit to the scaled \citet{chapman2005} 850\,\micron\ redshift
distribution. The SFRD figure, in particular, shows a large
discrepancy between the models, with the data significantly preferring
the $\alpha = 1$ model. To truly rule out one model or the other,
however, direct measurements of the redshift distribution are needed.

\begin{figure}
\centering
\includegraphics[width=\linewidth]{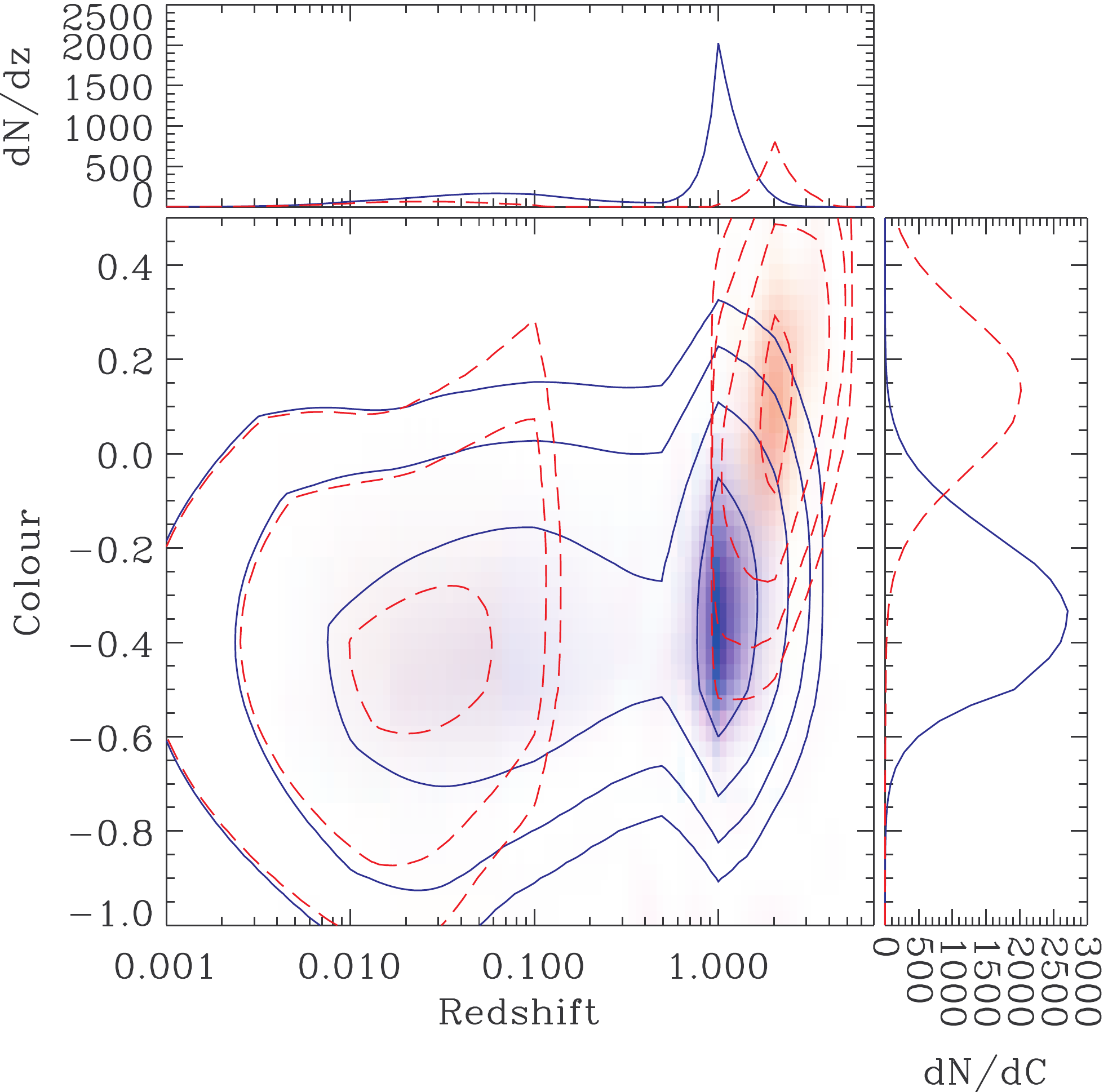}
\caption{The density of 350\,\micron\ sources brighter than 20\,mJy in
  the colour-redshift plane for both $\alpha=1$ (blue, solid contours)
  and $\alpha=0$ (red, long-dashed contours) models. The background
  intensity maps are linear and the contours are 1, 10, 100 and
  1000\,deg$^{-2}$ per unit colour per unit redshift. The
  distributions marginalised over redshift and colour are shown at the
  right and top, respectively.}
\label{fig:czplane}
\end{figure}

In Fig.~\ref{fig:czplane}, we show the integrated 350\,\micron\ counts
of sources brighter than 20\,mJy in the colour-redshift plane for both
models, with one-dimensional distributions shown in the side
panels. The low-redshift distributions are of course exactly the same,
but, as previously seen, the models are distinct at higher
redshifts. We see that the $\alpha=1$ model peaks at
$C\,{\sim}\,-0.4$, $z\,{\sim}\,1$, while the $\alpha=0$ model peaks at
$C\,{\sim}\,0.1$, $z\,{\sim}\,2$. A SPIRE redshift survey down to a
flux limit of 20\,mJy will allow us to unambiguously disentangle the
two models.

\begin{figure}
\centering
\includegraphics[width=\linewidth]{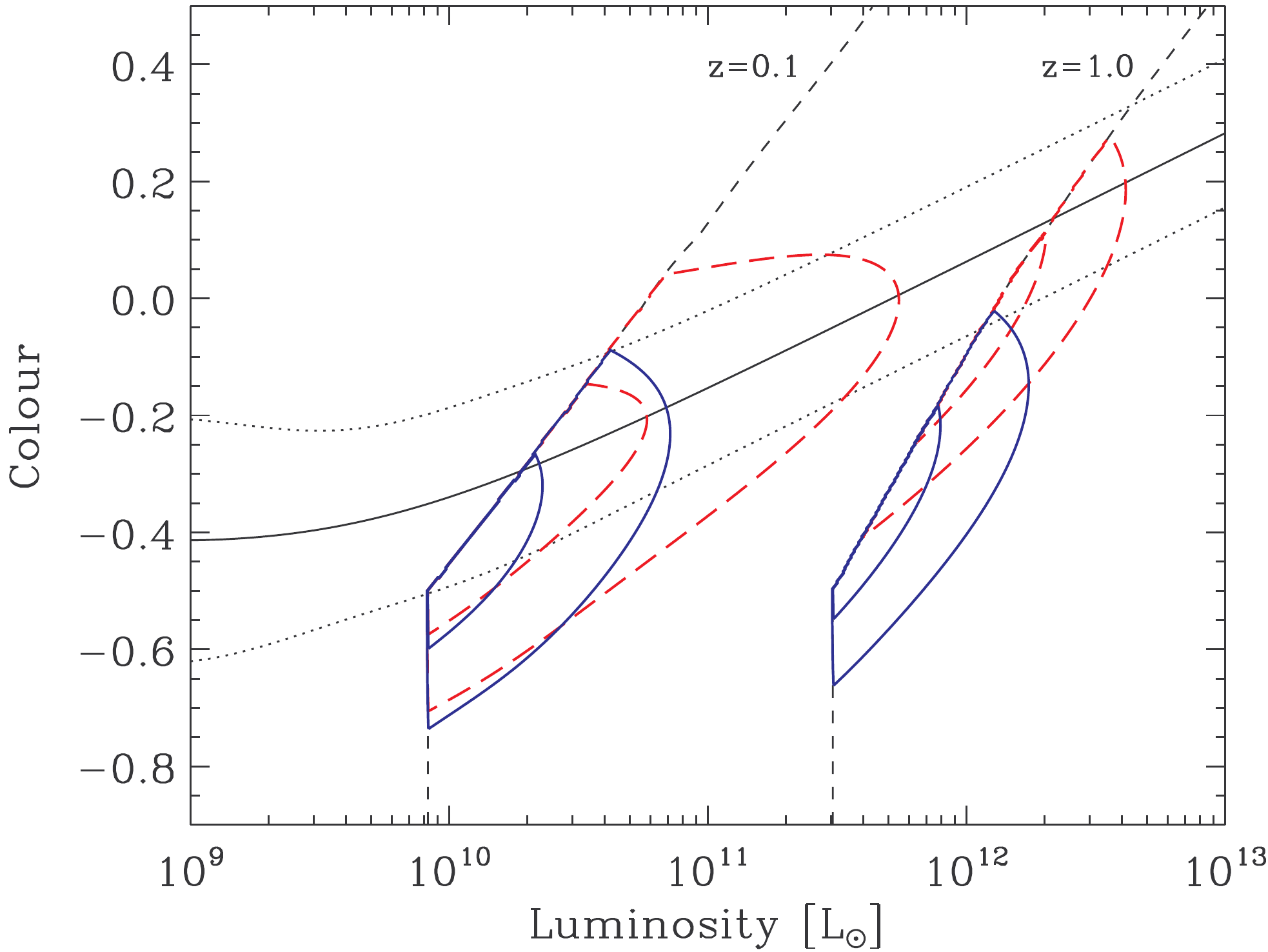}
\caption{The density of 350\,\micron\ sources brighter than 20\,mJy at
  a particular redshift in the colour-luminosity plane for both
  $\alpha=1$ (blue, solid contours) and $\alpha=0$ (red, long-dashed
  contours) models. We show the population at two redshifts, $z=0.1$
  and 1.0. The contours correspond to 0.1 and 0.5 of the maximum
  density. The thin short-dashed lines show the flux limits at the two
  redshifts. These lines are independent of luminosity below $C=-0.5$
  due to our choice of SED library. The solid black line shows the
  local colour-luminosity relationship, $p(C|L)$ (see
  Section~\ref{sec:locallc}), with 1-$\sigma$ limits plotted as dotted
  lines. This figure demonstrates the strong bias toward cooler and
  lower-luminosity sources when compared to the rest-frame
  distribution, caused by our selection function.}
\label{fig:clplane}
\end{figure}

When attempting to interpret the results of submm flux-limited
samples, we note the strong bias toward detecting cooler,
less-luminous sources. Following the prescription of
\citet{chapin2011}, we calculate the probability density in the
colour-luminosity plane at two fixed redshifts, $z=0.1$ and 1.0, for
samples selected at 350\,\micron\ above a flux density of 20\,mJy, for
both the $\alpha=1$ and $\alpha=0$ models. The results are shown in
Fig.~\ref{fig:clplane}. For reference, the locus of the
350\,\micron\ flux limit is shown as short-dashed lines at each
redshift, and the thin solid line shows the local colour-luminosity
correlation. The sample is biased downwards and to the left; this is
due to the fact that the total number of sources increases toward
lower luminosities, and that the selection function is
\textit{inclined\/} in the colour-luminosity plane; cooler sources may
be detected at fainter luminosities, where there are considerably more
objects. \citet{chapin2011} used this argument to suggest that their
submm-selected sample could be consistent with non-evolution of the
colour-luminosity correlation (despite finding a tail of cool sources
below the local colour-luminosity correlation). Similarly, included
among the Science Demonstration Phase results from \textsl{Herschel\/}
are some tentative indications that submm-selected sources are cooler
in the past \citep[e.g.,][]{hwang2010,elbaz2010}; it will be necessary
to model this selection effect to determine whether these observed
trends are indeed real.

Of course, another possibility is that the colour-luminosity
correlation evolves in a way completely different from the simple form
assumed in this paper. For example, \citet{symeonidis2010} argue that
the correlation simply broadens at high redshift, based on a sample of
galaxies spanning the submm--FIR, rather than there being a trend
toward lower temperatures in the past. While the observed correlation
at high-redshift may in fact be broader than the local correlation, we
note that the combination of a tight colour-luminosity correlation
with a submm selection effect can also broaden the apparent range of
colours, as shown in Fig.~\ref{fig:czplane}, and certainly
contributes, at least in part, to the effect seen by
\citet{symeonidis2010}.

Based on the forward-modelling approach taken in this paper, which
intrinsically accounts for these selection biases, we find that an
evolution toward cooler temperatures in the past is the most likely
scenario (i.e., the tail of cooler sources in the high-redshift
submm-selected population is \textit{even cooler\/} than what one
might expect given the selection biases). This conclusion is in rough
agreement with the model of \citet{valiante2009}, who also included
luminosity evolution in the colour-luminosity correlation to produce
cooler SEDs in the past, although they also included an extra
population of cold local sources that are presumably missed in
\textsl{IRAS\/} surveys. In contrast, \citet{lewis2005} and
\citet{leborgne2009} used fixed colour-luminosity correlations as a
function of redshift, and also appear to obtain reasonable results. We
note, however, that our result is based strongly on the imposed colour
distribution at the faint end of the luminosity function, which we
believe is possibly the cause of the over-predicted CIB. A modified
colour-luminosity relationship could have a large effect on the
relative merits of the $\alpha = 1$ and $\alpha = 0$ models. This is
discussed further in the next section.

As already discussed, the nature of the evolution of the
colour-luminosity correlation has a direct impact on our inferences
about the total SFRD as a function of time. As shown in
Fig.~\ref{fig:madau}, a decrease in the typical temperature of
galaxies at high redshift leads to a later peak formation epoch when
compared with a non-evolving colour-luminosity scenario. Working from
primarily SCUBA-selected samples, there is some evidence that
FIR-luminous galaxies at high-redshift are indeed cooler, but also
physically more extended, based on radio morphologies
\citep[e.g.,][]{chapman2004}, near--mid-IR colours
\citep[e.g.,][]{hainline2009}, and mid-IR spectra
\citep[e.g.,][]{menendez2009}. These observations appear to be
consistent with a local-Universe measurement that showed an
anti-correlation between physical size as a function of luminosity and
temperature \citep{chanial2007}.

\subsection{Reconciling the CIB with the SMG $z>1$ redshift
  distribution}
\label{sec:reconciling}

As has been seen in previous sections, the single-population model
that we have attempted to fit fails, primarily, in reconciling the
redshift distribution of $\sim$1\,mm-selected SMGs (and therefore the
$z>1$ SFRD, Figs.~\ref{fig:nzdist} and \ref{fig:madau}), with the
intensity of the CIB (Fig.~\ref{fig:cib}), despite doing a decent job
of fitting the counts in the bands that we have considered
(Figs.~\ref{fig:counts1} and \ref{fig:counts2}). Generally speaking,
we have found that models with lower (and therefore more consistent)
values of the CIB push the $\sim$1\,mm selected galaxies to lower (and
less consistent) redshifts.

We believe most of the discrepancy is due to our lack of knowledge of
the low-luminosity ($L < 10^{10}$\,L$_\odot$) galaxy SEDs. These faint
galaxies are essentially undetected individually above the confusion
limit in any of the existing blank-field ground-based $\sim$1\,mm
surveys \citep{blain2002}, nor in any BLAST
250--500\,\micron\ \citep{dye2009,chapin2011}, or \textsl{Spitzer\/}
70 and 160\,\micron\ surveys (see, for example, the description of the
COSMOS survey in Section~\ref{sec:colcol}). Even \textsl{IRAS\/}
struggled to conduct unbiased surveys of such faint objects, since it
was limited in sensitivity and also the nearby galaxies in the local
over-density were limited by sampling variance (see
Section~\ref{sec:locallc}). However, the integrated light from these
faint but numerous galaxies \emph{can\/} make a large contribution to
the CIB.

Referring to Fig.~4 of \chapin, it is clear that the clean correlation
between $C$ and $L$ broadens significantly between
$10^{10}$\,L$_\odot$ to $10^{9}$\,L$_\odot$, and then it completely
breaks down at even lower luminosities. Much of this effect is
probably due to the fact that the 60/100\,\micron\ flux ratio is
simply a poor indicator of the longer-wavelength SED of galaxies with
extremely cool dust (i.e., these bands sample relatively warmer and
potentially un-related dust). It is probably for this reason that
other authors have found it necessary to modify by hand the
distribution of low-luminosity / cool galaxies in their evolutionary
models.

In addition, the faint end of the FIR luminosity function contains a
significantly more heterogeneous collection of galaxies than the
bright end. Most luminous infrared galaxies ($L > 10^{11}$\,L$_\odot$)
and virtually all ultra-luminous infrared galaxies ($L >
10^{12}$\,L$_\odot$) are merger-driven starbursts
\citep{sanders1996}. In contrast, the faint end must include
everything else, including mildly star-forming spirals like the Milky
Way (relatively low-luminosity despite the presence of dust),
passively-evolving spheroids (extremely faint due to the near complete
lack of dust), but also smaller star-forming galaxies; blue compact
dwarfs, for example, have dust and contain a number of active
star-forming regions which may produce apparently large dust
temperatures despite their lower luminosities.

To test the hypothesis that our lack of knowledge at these low
luminosities plays a crucial role, we performed a simple test using
our best-fit $\alpha=1$ model. First, we assigned all galaxies below
$10^{10}$\,L$_\odot$ the \emph{warmest\/} SED (largest value of
$C$). This has the effect of \emph{decreasing\/} the flux densities at
$\lambda > 100$\,\micron, and \emph{increasing\/} the flux densities at
$\lambda < 60$\,\micron\ when extrapolating from the FIR luminosity
for those galaxies. We then re-calculated the number counts, CIB
spectrum, and redshift distribution of 850\,\micron\ selected
galaxies. While the modification to the SEDs of these faint galaxies
had little impact on either the counts or redshift distribution, where
data are available (again, since galaxies with such luminosities lie
below current confusion limits), there was a \emph{huge\/} change to the
CIB. Its predicted spectrum became much warmer (peaking closer to
100\,\micron\ than 200\,\micron), and dropped significantly below the
data at submm wavelengths. Conversely, assigning these low-luminosity
galaxies the coolest SED has the opposite effect, pushing the
predicted CIB peak to longer wavelengths, and exceeding the CIB by
even more than with the regular model.

We then repeated this procedure using a lower luminosity threshold of
$10^9$\,L$_\odot$. While the sense of the changes to the predicted
quantities were the same, the impact is obviously smaller, and using
the warmest SEDs for these faint galaxies, we obtain a reasonable fit
to the CIB (slightly low at 200\,\micron, and then going through the
data at longer submm wavelengths).

Based on this experiment, it is clear that how one treats
galaxies at the faint-end of the FIR luminosity function is crucial,
even although they are not well constrained observationally.  While one could
modify a phenomenological model \emph{ad hoc\/} to evolve the faint-end
significantly less than the bright-end as in \citet{lagache2003} and
\citet{valiante2009}, thus reducing the number of cool-galaxies at
higher redshifts, the alternative of increasing the temperatures of
fainter galaxies at \emph{all\/} redshifts would have a similar
effect. In the future one could use wide-area surveys such as
H-ATLAS \cite{eales2010} to measure the SEDs of fainter nearby galaxies,
and perhaps also with SCUBA-2 at
450\,\micron (since it will be capable of resolving most of the CIB
directly into individual sources; \citet{holland2006}).
We also note that since our present model tends to pull
down the SMG redshift distribution in order to improve the $\chi^2$ of
the CIB, our result that the $\alpha=1$ model is better than the
higher-redshift $\alpha=0$ model is far from secure. Soon it should be able to
perform tests such as Figs.~\ref{fig:czplane}--\ref{fig:clplane} to
determine, at least for the more luminous objects, whether there
really is evolution in the luminosity-temperature correlation using
SPIRE data, particularly from HerMES.

\subsection{Future Improvements}

Although in many ways our model is an improvement over previously
published studies, throughout the paper we have discussed several
shortcomings of the model and data sets used. Here we list some
analysis techniques and future data sets that will improve the quality
of the model:

\begin{itemize}
\item As discussed in Section~\ref{sec:rescounts}, the BLAST $P(D)$
  counts are inherently correlated. Instead of using the counts as an
  intermediate data set, we can use $P(D)$ \textit{within\/} the
  model-fitting framework to fit the maps directly. This is of course
  computationally intensive, but certainly worth pursuing. It also has
  the advantage over the \patanchon-style $P(D)$ in that the shape of
  the model is much more reflective of the shape of the true
  underlying counts, rather than imposing a connected power law onto
  the counts. Furthermore, as we noted in
  Section~\ref{sec:omitcounts}, the discrepancy between SCUBA
  850\,\micron\ and AzTEC 1.1\,mm counts may be due to the potentially
  biased counting of sources. $P(D)$ analyses do not suffer most of
  those problems.

\item New redshift distributions from SPIRE will go a long way to
  constraining the model, in particular breaking the colour-luminosity
  evolution degeneracy, and enabling more redshift nodes to be used.

\item Direct measurements of the local luminosity functions at FIR and
  submm wavelengths will give us a much better starting point for the
  evolving luminosity function, and will allow us to derive a better
  SED library. Wide-area PACS and SPIRE surveys will be able to
  provide these. We also hope that wide-area surveys will be
  capable of detecting and measuring the SEDs of galaxies with FIR
  luminosities in the range $10^{10}$--$10^{11}$\,L$_\odot$ to solve
  the discrepancy between the redshift distribution of SMGs and the
  spectrum of the CIB.

\item Extend the model to the mid-IR by including a more sophisticated
  SED library. This may require additional SED parameters to
  include, for example, the AGN contribution
  \citep[e.g.][]{valiante2009}.

\item Include lensing by adopting a similar approach to
  \citet{paciga2009}. Such a treatment is now necessary to explain the
  counts at mm wavelengths covering wide areas, and more recently,
  \textsl{Herschel\/}/SPIRE surveys.

\item More versatile modelling of the evolving SED distribution, in particular
  changing the width and shape of $P(C|L)$ as a function of $L$.

\end{itemize}

Finally, we note that a major part of the work in this paper went into
developing likelihood expressions for the various data sets. To
improve on our methodology for current and future surveys, it will be
important to fully characterise uncertainties and correlations between
data sets. For example, the differential number counts, integrated
background and redshift distributions of sources in the same field
will have a correlated cosmic variance term.

\section{Conclusions}
\label{sec:conclusions}

We have presented a sophisticated technique using MCMC to fit a simple
evolving luminosity function to a range of FIR and submm data. We are
able to measure errors on the evolutionary parameters and show the
correlations between these parameters. The results of the model are
available at \url{http://cmbr.phas.ubc.ca/model/}. We also show how
the various data sets are in tension with one another and demonstrate
the importance of redshift distributions.

An advantage of our approach over some other models in the literature
is that we need only consider the evolution of one galaxy population
with a single-parameter family of SEDs based on the correlation between
the 60-to-100\,\micron\ rest-frame colour and FIR luminosity. While we
find that, across the 70--1100\,\micron\ wavelength range, the
counts can be fit using models with and without evolution toward
cooler galaxy dust temperatures at higher-redshifts, there is
significant tension between the spectrum of the CIB and the redshift
distribution of SMGs. We believe that most of this discrepancy is
caused by the presently unknown distribution of submm SEDs for
galaxies with luminosities $<10^{10}$\,L$_\odot$. Emerging data from
\textsl{Herschel\/} will immediately help in two areas: (i) wide-area
surveys, such as H-ATLAS, should enable us to measure the SEDs for at
least a small sample of such objects; (ii) deeper surveys, such as
HerMES, can potentially be used to search for evolution in the
colour-luminosity correlation at the brighter end of the luminosity
function, as indicated in
Figs.~\ref{fig:czplane}--\ref{fig:clplane}. Such data should obviate
the need for \emph{ad hoc\/} modifications to the low-luminosity region
of the local luminosity function by, for example, using multiple
uncorrelated galaxy populations. The reality is that there is a
continuum of galaxy types, and it is our hope that we can gain more
realistic insight into how they form and evolve using the simplest
phenomenological models that are consistent with the data.

\section*{Acknowledgements}

We acknowledge the support of NASA through grant numbers NAG5-12785,
NAG5-13301, and NNGO-6GI11G, the NSF Office of Polar Programs, the
Canadian Space Agency, the Natural Sciences and Engineering Research
Council (NSERC) of Canada, and the UK Science and Technology
Facilities Council (STFC). We thank the authors of CosmoMC for
providing an easy-to-use MCMC generic sampler. Thanks also to Matthieu
B\'{e}thermin for useful discussions.

\bibliographystyle{mn2e}
\bibliography{mn-jour,refs}

\appendix

\label{lastpage}

\end{document}